\documentclass[10pt,journal]{IEEEtran}
\usepackage{cite}
\usepackage{graphicx}
\usepackage{amsmath}
\usepackage{times}
\usepackage{latexsym}
\usepackage{graphicx}
\usepackage{bm}
\usepackage{amssymb}
\usepackage[center]{caption2}
\usepackage{stfloats}
\usepackage{cases}
\usepackage{url}
\usepackage{array}
\usepackage{setspace}
\usepackage{fancyhdr}
\usepackage{color}
\usepackage{subfigure}
\usepackage{chemarrow}
\usepackage{extarrows}
\usepackage{algorithm}
\usepackage{algorithmic}

\newtheorem{theorem}{Theorem}

\newtheorem{corollary}{Corollary}
\newtheorem{proposition}{Proposition}

\newtheorem{remark}{Remark}

\def\proof{\noindent\hspace{2em}{\itshape Proof: }}
\def\endproof{\hspace*{\fill}~$\square$\par\endtrivlist\unskip}

\begin{document}
\title{Multipair Massive MIMO Relaying Systems with One-Bit ADCs and DACs}
\author{Chuili Kong, \emph{Student Member, IEEE}, Amine Mezghani, \emph{Member, IEEE}, Caijun Zhong, \emph{Senior Member, IEEE}, \\A. Lee Swindlehurst, \emph{Fellow, IEEE}, and Zhaoyang Zhang, \emph{Member, IEEE}
\thanks{C. Kong, C. Zhong and Z. Zhang are with the Institute of Information and Communication Engineering, Zhejiang University, Hangzhou 310027, China (e-mail: kcl\_dut@163.com; caijunzhong@zju.edu.cn; sunrise.heaven@gmail.com).}
\thanks{A. Mezghani and A. L. Swindlehurst are with the Center for Pervasive Communications and Computing, University of California, Irvine, CA 92697, USA (e-mail: amine.mezghani@uci.edu; swindle@uci.edu)}
\thanks{A. L. Swindlehurst and A. Mezghani were supported by
the National Science Foundation under Grant ECCS-1547155. A. L. Swindlehurst was further supported by the Technische Universit��at M��unchen Institute for Advanced Study, funded by the German Excellence Initiative and the European Union Seventh Framework Programme under grant agreement No. 291763, and by the European Union under the Marie Curie COFUND Program.
}
}
\maketitle
\begin{abstract}
This paper considers a multipair amplify-and-forward massive MIMO relaying system with one-bit ADCs and one-bit DACs at the relay. The channel state information is estimated via pilot training, and then utilized by the relay to perform simple maximum-ratio combining/maximum-ratio transmission processing. Leveraging on the Bussgang decomposition, an exact achievable rate is derived for the system with correlated quantization noise. Based on this, a closed-form asymptotic approximation for the achievable rate is presented, thereby enabling efficient evaluation of the impact of key parameters on the system performance. Furthermore, power scaling laws are characterized to study the potential energy efficiency associated with deploying massive one-bit antenna arrays at the relay. In addition, a power allocation strategy is designed to compensate for the rate degradation caused by the coarse quantization. Our results suggest that the quality of the channel estimates depends on the specific orthogonal pilot sequences that are used, contrary to unquantized systems where any set of orthogonal pilot sequences gives the same result. Moreover, the sum rate gap between the double-quantized relay system and an ideal non-quantized system is a moderate factor of $4/\pi^2$ in the low power regime.
\end{abstract}

\begin{keywords}
Massive MIMO, relays, one-bit quantization, power allocation
\end{keywords}
\section{Introduction}
Multipair multiple-input multiple-output (MIMO) relaying networks have recently attracted considerable attention since they can provide a cost-effective way of achieving performance gains in wireless systems via coverage extension and maintaining a uniform quality of service. In such a system, multiple sources simultaneously exchange information with multiple destinations via a shared multiple-antenna relay in the same time-frequency resource. Hence, multi-user interference is the primary system bottleneck. The deployment of massive antenna arrays at the relay has been proposed to address this issue due to their ability to suppress interference, provide large array and spatial multiplexing gains, and in turn to yield large improvements in spectral and energy efficiency \cite{H.Xie1,H.Xie3,C.Kong1,C.Kong2,M.Cheng}.

There has recently been considerable research interest in multipair massive MIMO relaying systems. For example, \cite{S.Jin} derived the ergodic rate of the system when maximum ratio combining/maximum ratio transmission (MRC/MRT) beamforming is employed and showed that the energy efficiency gain scales with the number of relay antennas in Rayleigh fading channels. Then, \cite{X.Wang} extended the analysis to the Ricean fading case and obtained similar power scaling behavior. For full-duplex systems, \cite{H.Q.Ngo,Z.Zhang} analytically compared the performance of MRC/MRT and zero-forcing reception/transmission and characterized the impact of the number of user pairs on the spectral efficiency.

All the aforementioned works are based on the assumption of perfect hardware. However, a large number of antennas at the relay implies a very large deployment cost and significant energy consumption if a separate RF chain is implemented for each antenna in order to maintain full beamforming flexibility.  In particular, the fabrication cost, chip area and power consumption of the analog-to-digital converters (ADCs) and the digital-to-analog converters (DACs) grow roughly exponentially with the number of quantization bits \cite{J.Yoo,R.H.Walden}. The cumulative cost and power required to implement a relay with a very large array can be prohibitive, and thus it is desirable to investigate the use of cheaper and more energy-efficient components, such as low-resolution (e.g., one bit) ADCs and DACs.  Fortunately, it has been shown in \cite{Y.Li3,Y.Li2} that large arrays exhibit a certain resilience to RF hardware impairments that could be caused by such low-cost components.
\subsection{Related Work}
Several recent contributions have investigated the impact of low-resolution ADCs on the massive MIMO uplink \cite{L.Fan,J.Zhang,L.Fan2,D.Verenzuela,Y.Li1,J.Choi2,C.Mollen,J.Mo,W.Tan,N.Liang}. For example, \cite{L.Fan2} optimized the training pilot length to maximize the spectral efficiency, while \cite{D.Verenzuela} revealed that in terms of overall energy efficiency, the optimal level of quantization is 4-5 bits. In \cite{Y.Li1}, the Bussgang decomposition \cite{J.J.Bussgang} was used to reformulate the nonlinear quantization using a second-order statistically equivalent linear operator, and to derive a linear minimum mean-squared error (LMMSE) channel estimator for one-bit ADCs. In \cite{J.Choi2}, a near-optimal low complexity bit allocation scheme was presented for millimeter wave channels exhibiting sparsity. The work of \cite{C.Mollen} examined the impact of one-bit ADCs on wideband channels with frequency-selective fading. Other work has focused on balancing the spectral and energy efficiency, either through the combined use of hybrid architectures with a small number of RF chains and low resolution ADCs, or using mixed ADCs architectures with high and low resolution.

In contrast to the uplink case, there are relatively fewer contributions that consider the massive MIMO downlink with low-resolution DACs. In \cite{S.Jacobsson}, it was shown that performance approaching the unquantized case can be achieved using DACs with only 3-4 bits of resolution. The nearly optimal quantized Wiener precoder with low-resolution DACs was studied in \cite{A.Mezghani4}, and the resulting solution was shown to outperform the conventional Wiener precoder with 4-6 bits of resolution at high signal-to-noise ratio (SNR). For the case of one-bit DACs, \cite{J.Guerreiro,Y.Li4} showed that even simple MRT precoding can achieve reasonable results. In \cite{O.B.Usman}, an LMMSE precoder was proposed by taking the quantization non-linearities into account, and different precoding schemes were compared in terms of uncoded bit error rate.
\subsection{Contributions}
All these prior works are for single-hop systems rather than dual-hop connections via a relay. Recently, \cite{J.Liu} considered a relay-based system that uses mixed-resolution ADCs at the base station. Unlike \cite{J.Liu}, we consider a multipair amplify-and-forward (AF) relaying system where the relay uses both one-bit ADCs and one-bit DACs. The one-bit ADCs cause errors in the channel estimation stage and subsequently in the reception of the uplink data; then, after a linear transformation, the one-bit DACs produce distortion when the downlink signal is coarsely quantized. In this paper, we present a detailed performance investigation of the achievable rate of such doubly quantized systems. In particular, the main contributions are summarized as follows:
\begin{itemize}
  \item We investigate a multipair AF relaying system that employs one-bit ADCs and DACs at the relay and uses MRC/MRT beamforming to process the signals. We take the correlation of the quantization noise into account, and present an exact achievable rate by using the arcsine law. Then, we use asymptotic arguments to provide an approximate closed-form expression for the achievable rate. Numerical results demonstrate that the approximate rate is accurate in typical massive MIMO scenarios, even with only a moderate number of users.
  \item We show that the channel estimation accuracy of the quantized system depends on the specific orthogonal pilot matrix that is used, which is in contrast to unquantized systems where any orthogonal pilot sequence yields the same result. We consider the specific case of identity and Hadamard pilot matrices, and we show that the identity training scheme provides better channel estimation performance for users with weaker than average channels, while the Hadamard training sequence is better for users with stronger channels.
  \item We compare the achievable rate of different ADC and DAC configurations, and show that a system with one-bit DACs and perfect ADCs outperforms a system with one-bit ADCs and perfect DACs. Focusing on the low transmit power regime, we show that the sum rate of the relay system with one-bit ADCs and DACs is $4/\pi^2$ times that achievable with perfect ADCs and DACs. Also, it is shown that the transmit power of each source or the relay can be reduced inversely proportional to the number of relay antennas, while maintaining a given quality-of-service.
  \item We formulate a power allocation problem to allocate power to each source and the relay, subject to a sum power budget. Locally optimum solutions are obtained by solving a sequence of geometric programming (GP) problems. Our numerical results suggest that the power allocation strategy can efficiently compensate for the rate degradation caused by the coarse quantization.
\end{itemize}
\subsection{Paper Outline and Notations}
The remainder of the paper is organized as follows. Section \ref{sec:system_model} introduces the multipair AF relaying system model under consideration. Section \ref{sec:rate_analysis} presents an approximate closed-form expression for the sum rate, and compares the rate achieved with different ADC and DAC configurations. Section \ref{sec:power_allocation} formulates a power allocation problem to compensate for the rate loss caused by the coarse quantization. Numerical results are provided in Section \ref{sec:numerical_results}. Finally, Section \ref{sec:conclusions} summarizes the key findings.

{\it Notation}: We use bold upper case letters to denote matrices, bold lower case letters to denote vectors and lower case letters to denote scalars. The notation $(\cdot)^{H}$, $(\cdot)^{*}$, $(\cdot)^{T}$, and $(\cdot)^{-1}$ respectively represent the conjugate transpose operator, the conjugate operator, the transpose operator, and the matrix inverse. The Euclidian norm is denoted by $|| \cdot ||$, the absolute value by $| \cdot |$, and ${\left[ {\bf{A}} \right]_{mn}}$ represents the $(m,n)$-th entry of $\bf{A}$. Also, ${\bf x} \thicksim {{\cal CN} ({\bf 0},{\bf \Sigma})}$ denote a circularly symmetric complex Gaussian random vector with zero mean and covariance matrix ${\bf \Sigma}$, while ${{\bf{I}}_k}$ is the identity matrix of size $k$. The symbol $\otimes$ is the Kronecker product, $\text{vec}\left(\bf A \right)$ represents a column vector containing the stacked columns of matrix $\bf A$, $\text{diag}\left(\bf B\right)$ denotes a diagonal matrix formed by the diagonal elements of matrix $\bf B$, $\Re\left( \bf C \right)$ and $\Im\left( \bf C  \right)$ stand for the real and imaginary part of $\bf C $, respectively. Finally, the statistical expectation operator is represented by ${\tt E}\{\cdot\}$, the variance operator is ${\text{Var}} \left(\cdot\right)$, and the trace is denoted by ${\text{tr}}\left(\cdot\right)$.
\section{System Model}\label{sec:system_model}
\begin{figure}[!ht]
    \centering
    \includegraphics[scale=0.22]{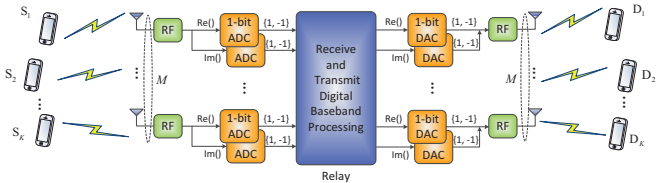}
    \caption{Illustration of the multipair half-duplex relaying with one-bit ADCs and DACs.}\label{fig:system_model}
  \end{figure}

Consider a multipair relaying system with one-bit quantization, as shown in Fig. \ref{fig:system_model}. There are $K$ single-antenna user pairs, denoted as ${\text S}_k$ and ${\text D}_{k}$, $k = 1,\ldots,K$, intending to exchange information with each other with the assistance of a shared relay. The relay is equipped with $M$ receive antennas with one-bit ADCs and $M$ transmit antennas with one-bit DACs. The one-bit ADCs cause errors in the channel estimation stage and subsequently in the reception of the uplink data; then, after a linear transformation, the one-bit DACs produce distortion when the downlink signal is coarsely quantized.  Thus, the system we study is double quantized. We assume that direct links between ${\text S}_{k}$ and ${\text D}_{k}$ do not exist due to large obstacles or severe shadowing. In addition, we further assume that the relay operates in half-duplex mode, and hence it cannot receive and transmit signals simultaneously. Accordingly, information transmission from ${\text S}_k$ to ${\text D}_{k}$ is completed in two phases. In the first phase, the $K$ sources transmit independent data symbols to the relay, and in the second phase the relay broadcasts the double-quantized signals ${\bf{\tilde x}}_\text R$ to the destinations. The signals at the relay's receive antennas and at the destinations  before quantization are respectively given by
\begin{align}\label{eq:yR}
  {\bf y}_{\text R} &= {\bf G}_\text{SR} {\bf P_\text S}^{1/2} {\bf x}_{\text S} + {\bf n}_{\text R}\\ \label{eq:yD}
  {\bf y}_\text D &= \gamma {\bf G}_\text{RD}^T {\bf{\tilde x}}_\text R + {\bf n}_\text D,
\end{align}
where $\gamma$ is chosen to satisfy a total power constraint $p_\text R$ at the relay, i.e., ${\tt E}\left\{||\gamma{\bf{\tilde x}}_\text R ||^2 \right\} = p_\text R$, which will be specified shortly. The source symbols are represented by ${\bf x}_{\text S} = [{ x}_{{\text S},1},\ldots,{ x}_{{\text S},K}]^T$, whose elements are assumed to be Gaussian distributed with zero mean and unit power. ${\bf P_\text S}$ is a diagonal matrix that denotes the transmit power of the sources with $\left[{\bf P}_\text S\right]_{kk} = p_{{\text S},k}$. The vectors ${\bf n}_{\text R}$ and ${\bf n}_\text D$ represent additive white Gaussian noise (AWGN) at the relay and destinations, whose elements are both identically and independently distributed (i.i.d.) ${\cal{CN}}(0,1)$. Note that to keep the notation clean and without loss of generality, we take the noise variance to be $1$ here, and also in the subsequent sections. With this convention, $p_\text S$ and also the subsequent transmit powers can be interpreted as the normalized SNR. The matrices ${\bf G}_\text{SR} = [{\bf g}_{{\text {SR}},1},\ldots,{\bf g}_{{\text {SR}},K}]$ and ${\bf G}_\text{RD} = [{\bf g}_{{\text {RD}},1},\ldots,{\bf g}_{{\text {RD}},K}]$ respectively represent the uncorrelated Rayleigh fading channels from the $K$ sources to the relay with ${\bf g}_{{\text {SR}},k} \in {\cal{CN}}\left(0,\beta_{\text{SR},k}{\bf I}_{M}\right)$ and the channels from the relay to the $K$ destinations with ${\bf g}_{{\text {RD}},k} \in {\cal{CN}}\left(0,\beta_{\text{RD},k}{\bf I}_{M}\right)$. The terms $\beta_{\text{SR},k}$ and $\beta_{\text{RD},k}$ model the large-scale path-loss, which is assumed to be constant over many coherence intervals and known a priori.
\subsection{Channel Estimation}
We assume training pilots are used to estimate the channel matrices ${\bf G}_\text{SR}$ and ${\bf G}_\text{RD}$, as in other massive MIMO AF relaying systems \cite{F.Gao}. Therefore, during each coherence interval of length $\tau_\text c$ (in symbols), all sources simultaneously transmit their mutually orthogonal pilot sequences ${\bf \Phi}_\text S \in \mathbb{C}^{{\tau_\text p} \times K}$ satisfying ${\bf \Phi}_\text S^H {\bf \Phi}_\text S = \tau_\text p {\bf I}_K$ to the relay while the destinations remain silent ($\tau_\text p \ge K$). Afterwards, all destinations simultaneously transmit their mutually orthogonal pilot sequences ${\bf \Phi}_\text D \in \mathbb{C}^{{\tau_\text p} \times K}$ satisfying ${\bf \Phi}_\text D^H {\bf \Phi}_\text D = \tau_\text p {\bf I}_K$ to the relay while the sources remain silent.

Since the channels ${\bf G}_\text{SR}$ and ${\bf G}_\text{RD}$ are estimated in the same fashion, we focus only on the first link ${\bf G}_\text{SR}$. The received training signal at the relay is given by
\begin{align}
  {\bf Y}_{\text p} = \sqrt{p_\text p} {\bf G}_\text{SR} {\bf \Phi}_\text S^T + {\bf N}_{\text p},
\end{align}
where $p_\text p$ represents the transmit power of each pilot symbol, and ${\bf N}_\text p$ denotes the noise at the relay, which has i.i.d. ${\cal{CN}}\left(0,1\right)$ elements. After vectorizing the matrix ${\bf Y}_{\text p}$, we obtain
\begin{align}
  {\bf y}_\text p = \text{vec} \left({\bf Y}_{\text p} \right) = \bar{\bf \Phi}_\text S \bar{\bf g}_\text{SR} + \bar{\bf n}_\text p,
\end{align}
where ${\bar{\bf \Phi}}_\text{S} = {\bf \Phi}_\text{S} \otimes \sqrt{p_\text p} {\bf I}_{M}$, ${\bar{\bf g}}_\text{SR} = \text{vec}\left({\bf G}_\text{SR}\right)$, and $\bar{\bf n}_\text p = \text{vec}\left({\bf N}_{\text p}\right)$.
\subsubsection{One-bit ADCs}
After the one-bit ADCs, the quantized signal can be expressed as
\begin{align}
  {\bf r}_\text p = {\cal Q}\left({\bf y}_\text p\right),
\end{align}
where ${\cal Q}\left(\cdot\right)$ denotes the one-bit quantization operation, which separately processes the real and imaginary parts of the signal. Therefore, the output set of the one-bit ADCs is $\frac{1}{\sqrt 2}\left\{\pm 1\pm 1j\right\}$. Using the Bussgang decomposition \cite{J.J.Bussgang,A.Mezghani}, ${\bf r}_{\text p}$ can be represented by a linear signal component and an uncorrelated quantization noise ${\bf q}_{\text p}$:
\begin{align}
  {\bf r}_\text p = {\bf A}_\text p {\bf y}_\text p + {\bf q}_\text p,
\end{align}
where ${\bf A}_\text p$ is the linear operator obtained by minimizing the power of the quantization noise ${\tt E}\left\{||{\bf q}_{\text p}||^2 \right\}$:
\begin{align}\label{eq:Aa}
  {\bf A}_{\text p} = {\bf R}_{{\bf y}_{\text p} {\bf{ r}}_{\text p}}^H {\bf R}_{{\bf y}_{\text p} {\bf y}_{\text p}}^{-1},
\end{align}
where ${\bf R}_{{\bf y}_{\text p} {\bf{r}}_{\text p}}$ denotes the cross-correlation matrix between the received signal ${\bf y}_{\text p}$ and the quantized signal ${\bf{r}}_{\text p}$, and ${\bf R}_{{\bf y}_{\text p} {\bf{y}}_{\text p}}$ represents the auto-correlation matrix of ${\bf y}_{\text p}$, which is computed as
\begin{align}\label{eq:R_yy}
  {\bf R}_{{\bf y}_{\text p} {\bf y}_{\text p}} =  {\bar{\bf \Phi}}_\text S {\tilde{\bf D}}_\text{SR} {\bar{\bf \Phi}}_\text S^H + {\bf I}_{M \tau_\text p},
\end{align}
where ${\tilde{\bf D}}_\text{SR} = \left({\bf D}_\text{SR} \otimes {\bf I}_{M}\right)$ and ${\bf D}_\text{SR}$ is a diagonal matrix whose elements are $\left[{\bf D}_\text{SR}\right]_{kk} = \beta_{\text{SR},k}$ for $k = 1,\ldots, K$.

For one-bit quantization, by invoking the results in \cite[Chapter 10]{A.Papoulis} and applying the arcsine law \cite{G.Jacovitti}, we have
\begin{align}\label{eq:Ryy_tilde}
  {\bf R}_{{\bf y}_{\text p} {\bf{r}}_{\text p}} &= \frac{2}{\pi} {\bf R}_{{\bf y}_{\text p} {\bf y}_{\text p}} {\text{diag}} \left( {\bf R}_{{\bf y}_{\text p} {\bf y}_{\text p}} \right)^{-1/2}\\
  {\bf R}_{{\bf{r}}_{\text p} {\bf{r}}_{\text p}} &= \frac{2}{\pi} \left( {\text{arcsin}} \left( {\bf J}\right) + j {\text{arcsin}} \left( {\bf K} \right) \right), \label{eq:Ry_tilde}
\end{align}
where
\begin{align}
{\bf J} &= {\text{diag}} \left( {\bf R}_{{\bf y}_{\text p} {\bf y}_{\text p}} \right)^{-1/2} \Re\left( {\bf R}_{{\bf y}_{\text p} {\bf y}_{\text p}} \right) {\text{diag}} \left( {\bf R}_{{\bf y}_{\text p} {\bf y}_{\text p}} \right)^{-1/2} \\
{\bf K} &= {\text{diag}} \left( {\bf R}_{{\bf y}_{\text p} {\bf y}_{\text p}} \right)^{-1/2} \Im \left( {\bf R}_{{\bf y}_{\text p} {\bf y}_{\text p}} \right) {\text{diag}} \left( {\bf R}_{{\bf y}_{\text p} {\bf y}_{\text p}} \right)^{-1/2}.
\end{align}
Substituting \eqref{eq:Ryy_tilde} into \eqref{eq:Aa}, and after some simple mathematical manipulations, we have
\begin{align}\label{eq:Ap:diag}
  {\bf A}_{\text p} = \sqrt{\frac{2}{\pi}} {\text{diag}} \left( {\bf R}_{{\bf y}_{\text p} {\bf y}_{\text p}} \right)^{-1/2}.
\end{align}

Since ${\bf q}_{\text p}$ is uncorrelated with ${\bf y}_{\text p}$, we have
\begin{align}\label{eq:R_qp}
 {\bf R}_{{\bf q}_{\text p} {\bf q}_{\text p}} = {\bf R}_{{\bf{r}}_{\text p} {\bf{r}}_{\text p}} - {\bf A}_{\text p} {\bf R}_{{\bf y}_{\text p} {\bf{y}}_{\text p}} {\bf A}_{\text p}^H.
 \end{align}
Substituting \eqref{eq:Ry_tilde} into \eqref{eq:R_qp} yields
 \begin{align}\label{eq:R_qp:arcsine}
   {\bf R}_{{\bf q}_{\text p} {\bf q}_{\text p}} = \frac{2}{\pi} \left( {\text{arcsin}} \left( {\bf J}\right) + j {\text{arcsin}} \left( {\bf K} \right) \right) - \frac{2}{\pi} \left({\bf J} + j {\bf K} \right).
 \end{align}
 \subsubsection{LMMSE estimator}
 Based on the observation ${\bf r}_\text p$ and the training pilots ${\bf \Phi}_\text S$, we use the LMMSE technique to estimate ${\bf G}_\text{SR}$. Hence, the estimated channel ${\hat{\bf g}}_\text{SR}$ is given by
 \begin{align}
   {\hat{\bf g}}_\text{SR} = {\bf R}_{{\bar{\bf g}}_\text{SR} {\bf r}_\text p} {\bf R}_{{\bf r}_\text p {\bf r}_\text p}^{-1} {\bf r}_\text p.
 \end{align}
As a result, the covariance matrix of the estimated channel ${\hat{\bf g}}_\text{SR}$ is expressed as
 \begin{align}\label{eq:R:g_hat_SR}
   &{\bf R}_{{\hat{\bf g}}_\text{SR} {\hat{\bf g}}_\text{SR}} = \\ \notag
   &{\tilde{\bf D}}_\text{SR} {\tilde{\bf \Phi}}_\text S^H \left({\tilde{\bf \Phi}}_\text S {\tilde{\bf D}}_\text{SR} {\tilde{\bf \Phi}}_\text S^H + {\bf A}_\text p {\bf A}_\text p^H + {\bf R}_{{\bf q}_{\text p} {\bf q}_{\text p}} \right)^{-1} {\tilde{\bf \Phi}}_\text S {\tilde{\bf D}}_\text{SR},
 \end{align}
 where ${\tilde{\bf \Phi}}_\text S = {\bf A}_\text p {\bar{\bf \Phi}}_\text S$.

 \begin{remark}
   From \eqref{eq:R:g_hat_SR}, we can see that ${\bf R}_{{\hat{\bf g}}_\text{SR} {\hat{\bf g}}_\text{SR}}$ is a non-trivial function of ${\tilde{\bf \Phi}}_\text S$, which indicates that the quality of the channel estimates depends on the specific realization of the pilot sequence, which is contrary to unquantized systems where any set of orthogonal pilot sequences gives the same result.
 \end{remark}
 \begin{remark}
 Although our conclusion in {\em Remark 1} is obtained based on the LMMSE estimator, it also holds for the maximum likelihood estimator \cite{M.T.Ivrlac}.
 \end{remark}

  In the following, we study the performance of two specific pilot sequences to show how the pilot matrix affects the channel estimation. Here, we choose $\tau_\text p = K$, which is the minimum possible length of the pilot sequence.

 a) \textit{\underline{Identity Matrix}}. In this case, ${\bf{\Phi}}_\text S = \sqrt{K} {\bf I}_K$, and hence we have
 \begin{align}
   {\bf R}_{{\bf y}_{\text p} {\bf y}_{\text p}} =  K p_\text p {\tilde{\bf D}}_\text{SR}  + {\bf I}_{M K}.
 \end{align}
Consequently,
 \begin{align}\label{eq:Ap:case1}
   {\bf A}_{\text p} &=  \sqrt{\frac{2}{\pi}} \left( K p_\text p {\tilde{\bf D}}_\text{SR}  + {\bf I}_{M K} \right)^{-1/2} = {\bar{\bf A}}_\text p \otimes {\bf I}_{M}\\
   {\bf R}_{{\bf q}_{\text p} {\bf q}_{\text p}} &= \left(1 - \frac{2}{\pi} \right) {\bf I}_{M K}, \label{eq:Rqp:case1}
 \end{align}
where ${\bar{\bf A}}_\text p$ is a diagonal matrix with $\left[{\bar{\bf A}}_\text p\right]_{kk} = \alpha_{\text p,k}= \sqrt{\frac{2}{\pi} \frac{1}{K p_\text p \beta_{\text{SR},k} + 1}}$. Substituting \eqref{eq:Ap:case1} and \eqref{eq:Rqp:case1} into \eqref{eq:R:g_hat_SR}, we obtain
\begin{align}
{\bf R}_{{\hat{\bf g}}_\text{SR} {\hat{\bf g}}_\text{SR}} = {\bf Q}_\text{SR}^{\left(1\right)} \otimes {\bf I}_{M},
\end{align}
where ${\bf Q}_\text{SR}^{\left(1\right)}$ is a diagonal matrix with elements
\begin{align}\label{eq:sigma_SR}
  \left[{\bf Q}_\text{SR}^{\left(1\right)} \right]_{kk} = \sigma_{\text{SR},k}^2 = \frac{2}{\pi} \frac{K p_\text p \beta_{\text{SR},k}^2}{K p_\text p \beta_{\text{SR},k} + 1}.
\end{align}

 b) \textit{\underline{Hadamard Matrix}}. In this case, every element of ${\bf{\Phi}}_\text S$ is $+1$ or $-1$, and hence we have
\begin{align}\label{eq:Ap:case2}
  {\bf A}_\text p &= \sqrt{\frac{2}{\pi} \frac{1}{p_\text p \sum\limits_{n=1}^K \beta_{\text{SR},k} + 1}} {\bf I}_{M K} \\
  {\bf R}_{{\bf q}_{\text p} {\bf q}_{\text p}} &\approx \left(1 - \frac{2}{\pi} \right) {\bf I}_{M K}, \label{eq:Rqp:case2}
\end{align}
where the approximation in \eqref{eq:Rqp:case2} holds for low $p_\text p$. Substituting \eqref{eq:Ap:case2} and \eqref{eq:Rqp:case2} into \eqref{eq:R:g_hat_SR}, we obtain
\begin{align}
{\bf R}_{{\hat{\bf g}}_\text{SR} {\hat{\bf g}}_\text{SR}} = {\bf Q}_\text{SR}^{\left(2\right)} \otimes {\bf I}_{M},
\end{align}
where ${\bf Q}_\text{SR}^{\left(2\right)}$ is a diagonal matrix with entries
\begin{align}
\left[{\bf Q}_\text{SR}^{\left(2\right)} \right]_{kk} = \kappa_{\text{SR},k}^2 = \frac{K {\bar\alpha}_{\text p}^2 \beta_{\text{SR},k}^2 p_\text p}{K {\bar\alpha}_{\text p}^2 \beta_{\text{SR},k} p_\text p + {\bar\alpha}_{\text p}^2 + 1 - \frac{2}{\pi}},
\end{align}
 where
 \begin{align}
 {\bar\alpha}_\text p = \sqrt{\frac{2}{\pi}\frac{1}{p_\text p \sum\limits_{k=1}^K\beta_{\text{SR},k} + 1}}.
 \end{align}

For both cases, the channels from the sources to the relay ${\bf g}_{\text{SR},k}$ can be decomposed as
\begin{align}
  {\bf g}_{\text{SR},k} = {\bf{\hat g}}_{\text{SR},k} + {\bf e}_{\text{SR},k},
\end{align}
where ${\bf e}_{\text{SR},k}$ is the estimation error vector. The elements of ${\bf{\hat g}}_{\text{SR},k}$ and ${\bf e}_{\text{SR},k}$ are respectively distributed as ${\cal{CN}}(0, \sigma_{\text{SR},k}^2 )$ and ${\cal{CN}}(0, {\tilde\sigma}_{\text{SR},k}^2 )$ when ${\bf \Phi}_\text{SR}$ is an identity matrix, while they are distributed as ${\cal{CN}}(0, \kappa_{\text{SR},k}^2 )$ and ${\cal{CN}}(0, {\tilde\kappa}_{\text{SR},k}^2)$ when ${\bf \Phi}_\text{SR}$ is a Hadamard matrix, where ${\tilde\sigma}_{\text{SR},k}^2 = \beta_{\text{SR},k} - \sigma_{\text{SR},k}^2$ and ${\tilde\kappa}_{\text{SR},k}^2 = \beta_{\text{SR},k} - \kappa_{\text{SR},k}^2$. In what follows we define ${\bf{\hat G}}_\text{SR} = [{\bf{\hat g}}_{{\text {SR}},1},\ldots,{\bf{\hat g}}_{{\text {SR}},K}]$ and ${\bf E}_\text{SR} = [{\bf e}_{{\text {SR}},1},\ldots,{\bf e}_{{\text {SR}},K}]$.

Similarly, the channels from the relay to the destinations ${\bf g}_{\text{RD},k}$ can be decomposed as
\begin{align}
  {\bf g}_{\text{RD},k} = {\bf{\hat g}}_{\text{RD},k} + {\bf e}_{\text{RD},k},
\end{align}
where ${\bf{\hat g}}_{\text{RD},k}$ and ${\bf e}_{\text{RD},k}$ are the estimated channel and estimation error vectors. The elements of ${\bf{\hat g}}_{\text{RD},k}$ and ${\bf e}_{\text{RD},k}$ are distributed as ${\cal{CN}}(0, \sigma_{\text{RD},k}^2 )$ and ${\cal{CN}}(0, {\tilde\sigma}_{\text{RD},k}^2 )$ when ${\bf \Phi}_\text{RD}$ is an identity matrix, while they are ${\cal{CN}}(0, \kappa_{\text{RD},k}^2 )$ and ${\cal{CN}}(0, {\tilde\kappa}_{\text{RD},k}^2 )$ when ${\bf \Phi}_\text{RD}$ is a Hadamard matrix, where
\begin{align}\label{eq:sigma_RD}
\sigma_{\text{RD},k}^2 &=  \frac{2}{\pi} \frac{K p_\text p \beta_{\text{RD},k}^2}{K p_\text p \beta_{\text{RD},k} + 1}\\
\kappa_{\text{RD},k}^2 &= \frac{K {\hat\alpha}_{\text p}^2 \beta_{\text{RD},k}^2 p_\text p}{K {\hat\alpha}_{\text p}^2 \beta_{\text{RD},k} p_\text p + {\hat\alpha}_{\text p}^2 + 1 - \frac{2}{\pi}},
 \end{align}
  with
   \begin{align}
   {\hat\alpha}_\text p = \sqrt{\frac{2}{\pi}\frac{1}{p_\text p \sum\limits_{k=1}^K\beta_{\text{RD},k} + 1}},
   \end{align}
   and ${\tilde\sigma}_{\text{RD},k}^2 = \beta_{\text{RD},k} - \sigma_{\text{RD},k}^2$, ${\tilde\kappa}_{\text{RD},k}^2 = \beta_{\text{RD},k} - \kappa_{\text{RD},k}^2$. We also define ${\bf{\hat G}}_\text{RD} = [{\bf{\hat g}}_{{\text {RD}},1},\ldots,{\bf{\hat g}}_{{\text {RD}},K}]$ and ${\bf E}_\text{RD} = [{\bf e}_{{\text {RD}},1},\ldots,{\bf e}_{{\text {RD}},K}]$.

For the channel from the \emph{k}-th source to the relay, the mean-square error (MSE) is given by
\begin{align}
  \text{MSE}_{\text{SR},k} = {\tt E}\left\{ ||{\bf{\hat g}}_{\text{SR},k} - {\bf{ g}}_{\text{SR},k}||^2 \right\}.
\end{align}
Based on the above results, we have $\text{MSE}_{\text{SR},k} = {\tilde\sigma}_{\text{SR},k}^2$ for the identity matrix and $\text{MSE}_{\text{SR},k} = {\tilde\kappa}_{\text{SR},k}^2$ for the Hadamard matrix. The following proposition compares the MSE of the two approaches.
\begin{proposition}\label{theor:pilot_matrix}
  For estimating the channel ${\bf g}_{\text{SR},k}$, the identity matrix is preferable to the Hadamard matrix for user $k$ if $\beta_{\text{SR},k} < \frac{1}{K} \sum\limits_{i=1}^K \beta_{\text{SR},i}$, and vice versa.
\end{proposition}
\proof The proof is trivial since ${\tilde\sigma}_{\text{SR},k}^2 < {\tilde\kappa}_{\text{SR},k}^2$ if $\beta_{\text{SR},k} < \frac{1}{K} \sum\limits_{i=1}^K \beta_{\text{SR},i}$. \endproof

Proposition \ref{theor:pilot_matrix} reveals that the accuracy of the individual channel estimates depends on the particular choice of the orthogonal training scheme, contrary to the ideal case without quantization. More precisely, the
scaled identity matrix is beneficial for any user with higher path loss
than the average. This is because a weak user benefits from being the only one transmitting at a given time, without the presence of stronger users that dominate the behavior of the ADC. In the case of  Hadamard matrix, all users are
transmitting simultaneously, resulting in an average quantization noise
level for all users jointly, which is advantageous  for users with
stronger channels.

The question of optimizing the pilot sequence for a given performance metric is an interesting one, but is beyond the scope of the paper. For simplicity, we will assume the identity matrix approach in which each user's channel is estimated one at a time.
\subsection{Data Transmission}
\subsubsection{Quantization with One-bit ADCs}
With one-bit ADCs at the receiver, the resulting quantized signals can be expressed as
\begin{align}\label{eq:yR_tilde}
  {\bf{\tilde y}}_{\text R} = {\cal Q}\left( {\bf y}_{\text R} \right) = {\bf A}_{\text a} {\bf y}_{\text R} + {\bf q}_{\text a},
\end{align}
where ${\bf A}_\text a$ is the linear operator, which is uncorrelated with ${\bf y}_\text R$. By adopting the same technique in the previous subsection, we have
\begin{align}\label{eq:Aa:diag}
  {\bf A}_{\text a} &= \sqrt{\frac{2}{\pi}} {\text{diag}} \left( {\bf R}_{{\bf y}_{\text R} {\bf y}_{\text R}} \right)^{-1/2}\\
  {\bf R}_{{\bf q}_{\text a} {\bf q}_{\text a}} &= \frac{2}{\pi} \left( {\text{arcsin}} \left( {\bf X}\right) + j {\text{arcsin}} \left( {\bf Y} \right) \right) - \frac{2}{\pi} \left({\bf X} + j {\bf Y} \right), \label{eq:R_qa:arcsine}
\end{align}
where
\begin{align} \notag
{\bf X} &= {\text{diag}} \left( {\bf R}_{{\bf y}_{\text R} {\bf y}_{\text R}} \right)^{-1/2} \Re\left( {\bf R}_{{\bf y}_{\text R} {\bf y}_{\text R}} \right) {\text{diag}} \left( {\bf R}_{{\bf y}_{\text R} {\bf y}_{\text R}} \right)^{-1/2} \\ \notag
{\bf Y} &= {\text{diag}} \left( {\bf R}_{{\bf y}_{\text R} {\bf y}_{\text R}} \right)^{-1/2} \Im \left( {\bf R}_{{\bf y}_{\text R} {\bf y}_{\text R}} \right) {\text{diag}} \left( {\bf R}_{{\bf y}_{\text R} {\bf y}_{\text R}} \right)^{-1/2}\\ \notag
{\bf R}_{{\bf y}_{\text R} {\bf y}_{\text R}} &=  {\bf G}_\text{SR} {\bf P}_\text S {\bf G}_\text{SR}^H + {\bf I}_{M}.
\end{align}
\subsubsection{Digital Linear Processing}
We assume that the relay adopts an AF protocol to process the quantized signals by one-bit ADCs ${\tilde{\bf y}}_\text R$, yielding
\begin{align}\label{eq:xR}
  {\bf x}_\text R = {\bf W} {\bf {\tilde y}}_\text R,
\end{align}
where ${\bf W} = {\hat{\bf G}}_\text{RD}^* {\hat{\bf G}}_\text{SR}^H$ for MRC/MRT beamforming.
\subsubsection{Quantization with One-bit DACs}
Assuming one-bit DACs at the transmitter, the resulting quantized signals to be sent by the relay's transmit antennas can be expressed as
\begin{align}\label{eq:x_tilde}
  {\bf{\tilde x}}_\text R = {\cal Q} \left({\bf x}_\text R\right) = {\bf A}_\text d {\bf x}_\text R + {\bf q}_\text d,
\end{align}
where ${\bf A}_\text d$ is the linear operator, and ${\bf q}_\text d$ is the quantization noise at the relay's transmit antennas, which is uncorrelated with ${\bf x}_\text R$. Due to the one-bit DACs, we have ${\tt E}\left\{||{\bf{\tilde x}}_\text R ||^2 \right\} = M$. Therefore, the normalization factor $\gamma$ (c.f. \eqref{eq:yD}) can be expressed as
\begin{align}\label{eq:gamma}
  \gamma = \sqrt{\frac{p_\text R}{M}}.
\end{align}

Following in the same fashion as with the ADCs derivations, we obtain
\begin{align}\label{eq:Ad}
  {\bf A}_\text d &= \sqrt{\frac{2}{\pi}} {\text{diag}} \left( {\bf R}_{{\bf x}_{\text R} {\bf x}_{\text R}} \right)^{-1/2}\\ \label{eq:R_qd}
  {\bf R}_{{\bf q}_{\text d} {\bf q}_{\text d}} &= \frac{2}{\pi} \left( {\text{arcsin}} \left( {\bf U}\right) + j {\text{arcsin}} \left( {\bf V} \right) \right) - \frac{2}{\pi} \left({\bf U} + j {\bf V} \right),
\end{align}
where
\begin{align} \notag
{\bf U} &= {\text{diag}} \left( {\bf R}_{{\bf x}_{\text R} {\bf x}_{\text R}} \right)^{-1/2} \Re\left( {\bf R}_{{\bf x}_{\text R} {\bf x}_{\text R}} \right) {\text{diag}} \left( {\bf R}_{{\bf x}_{\text R} {\bf x}_{\text R}} \right)^{-1/2} \\ \notag
{\bf V} &= {\text{diag}} \left( {\bf R}_{{\bf x}_{\text R} {\bf x}_{\text R}} \right)^{-1/2} \Im \left( {\bf R}_{{\bf x}_{\text R} {\bf x}_{\text R}} \right) {\text{diag}} \left( {\bf R}_{{\bf x}_{\text R} {\bf x}_{\text R}} \right)^{-1/2}\\ \notag
  {\bf R}_{{\bf x}_{\text R} {\bf x}_{\text R}} &= {\bf W}  {\bf R}_{{\bf{\tilde y}}_{\text R} {\bf{\tilde y}}_{\text R}} {\bf W}^H \\ \notag
    {\bf R}_{{\bf{\tilde y}}_{\text R} {\bf{\tilde y}}_{\text R}} &= {\bf A}_\text a {\bf R}_{{\bf{ y}}_{\text R} {\bf{ y}}_{\text R}} {\bf A}_\text a^H + {\bf R}_{{\bf q}_{\text a} {\bf q}_{\text a}}.
\end{align}
\section{Achievable Rate Analysis}\label{sec:rate_analysis}
In this section, we investigate the achievable rate of the considered system. In particular, we first provide an expression for the exact achievable rate, which is applicable to arbitrary system configurations. Then we use asymptotic arguments to derive an approximate rate to provide some key insights.
\subsection{Exact Achievable Rate Analysis}
We consider the realistic case where the $K$ destinations do not have access to the instantaneous CSI, which is a typical assumption in the massive MIMO literature since the dissemination of instantaneous CSI leads to excessively high computational and signaling costs for very large antenna arrays. Hence, $\text D_k$ uses only statistical CSI to decode the desired signal. Combining \eqref{eq:yR}, \eqref{eq:yD}, \eqref{eq:yR_tilde}, \eqref{eq:xR}, \eqref{eq:x_tilde}, and \eqref{eq:gamma} yields the received signal at the \emph{k}-th destination

\begin{align}\label{eq:received_signal_exact}
  y_{\text D,k} &= \underbrace{\gamma \sqrt{p_{{\text S},k}} {\tt E} \left\{ {\bf g}_{\text{RD},k}^T {\bf A}_\text d {\bf W} {\bf A}_\text a {\bf g}_{\text{SR},k} \right\} x_{\text S,k}}_\text{desired signal} + \underbrace{{\tilde n}_{\text D,k}}_\text{effective noise},
\end{align}
where
where ${\tilde n}_{\text D,k} = \\ \underbrace{\gamma \sqrt{p_{{\text S},k}} \left( {\bf g}_{\text{RD},k}^T {\bf A}_\text d {\bf W} {\bf A}_\text a {\bf g}_{\text{SR},k} - {\tt E} \left\{ {\bf g}_{\text{RD},k}^T {\bf A}_\text d {\bf W} {\bf A}_\text a {\bf g}_{\text{SR},k} \right\} \right) x_{\text S,k}}_\text{estimation error} \\
  + \underbrace{\gamma \sum\limits_{i \neq k} \sqrt{p_{{\text S},i}} {\bf g}_{\text{RD},k}^T {\bf A}_\text d {\bf W} {\bf A}_\text a {\bf g}_{\text{SR},i} x_{\text S,i}}_\text{interpair interference} + \underbrace{\gamma  {\bf g}_{\text{RD},k}^T {\bf A}_\text d {\bf W} {\bf A}_\text a {\bf n}_\text R}_\text{noise at the relay} \\ + \underbrace{\gamma {\bf g}_{\text{RD},k}^T {\bf A}_\text d {\bf W} {\bf q}_\text a}_\text{quantization noise of ADCs}
  + \underbrace{ \gamma {\bf g}_{\text{RD},k}^T {\bf q}_\text d}_\text{quantization noise of DACs} + \underbrace{n_{\text D,k}}_\text{noise at $k$-th destination}$,
where $n_{\text D,k}$ is the \emph{k}-th element of the noise vector ${\bf n}_\text D$. Noticing that the ``desired signal'' and the ``effective noise'' in \eqref{eq:received_signal_exact} are uncorrelated, and capitalizing on the fact that the worst-case uncorrelated additive noise is independent Gaussian, we obtain the following achievable rate for the \emph{k}-th destination:
\begin{align}\label{eq:rate_exact}
   &R_k= \\
   &\frac{\tau_\text c - 2 \tau_\text p}{2 \tau_\text c} \log_2\left(1 + \frac{A_k}{B_k + C_k + D_k + E_k + F_k + \frac{1}{\gamma^2}} \right), \notag
\end{align}
where
\begin{align}
A_k &= p_{{\text S},k} |{\tt E}\left\{ {\bf g}_{\text{RD},k}^T {\bf{ A}}_\text d {\bf W} {\bf{A}}_\text a {\bf g}_{\text{SR},k}\right\} |^2\\
B_k &= p_{{\text S},k} \text{Var}\left( {\bf g}_{\text{RD},k}^T {\bf{ A}}_\text d {\bf W} {\bf{A}}_\text a {\bf g}_{\text{SR},k} \right)\\
C_k &= \sum\limits_{i\neq k} p_{{\text S},i} {\tt E} \left\{ |{\bf g}_{\text{RD},k}^T {\bf{ A}}_\text d {\bf W} {\bf{ A}}_\text a {\bf g}_{\text{SR},i}|^2 \right\}\\
D_k &= {\tt E}\left\{ ||{\bf g}_{\text{RD},k}^T {\bf{A}}_\text d {\bf W} {\bf{A}}_\text a ||^2\right\}\\
E_k &= {\tt E}\left\{|{\bf g}_{\text{RD},k}^T {\bf{ A}}_\text d {\bf W} {\bf R}_{{\bf q}_{\text a} {\bf q}_{\text a}} {\bf W}^H {\bf{ A}}_\text d^H {\bf g}_{\text{RD},k}^* |\right\}\\
F_k &= {\tt E}\left\{|{\bf g}_{\text{RD},k}^T {\bf R}_{{\bf q}_{\text d} {\bf q}_{\text d}} {\bf g}_{\text{RD},k}^* |\right\}.
\end{align}

\subsection{Asymptotic Simplifications}
 As we can see, the matrices ${\bf R}_{{\bf q}_{\text a} {\bf q}_{\text a}}$, ${\bf{ A}}_\text d$, and ${\bf R}_{{\bf q}_{\text d} {\bf q}_{\text d}}$ all involve arcsine functions, which does not give much insight into how the rate changes with various parameters. To facilitate the analysis, we focus on the asymptotic regime for a large number of users, in which \eqref{eq:R_yy} can be approximated by

\begin{align}\label{eq:R_yy:approx}
  {\bf R}_{{\bf y}_{\text R} {\bf y}_{\text R}} \approx {\text{diag}} \left( {\bf R}_{{\bf y}_{\text R} {\bf y}_{\text R}} \right) \approx \left(1 + \sum\limits_{k=1}^K p_{{\text S},k} \beta_{\text{SR},k} \right) {\bf I}_{M}.
\end{align}
Substituting \eqref{eq:R_yy:approx} into \eqref{eq:Aa:diag} and \eqref{eq:R_qa:arcsine}, we have
\begin{align}\label{eq:Aa:approx}
  {\bf A}_{\text a} &\approx \sqrt{\frac{2}{\pi}} \sqrt{\frac{1}{1 + \sum\limits_{k=1}^K p_{{\text S},k} \beta_{\text{SR},k}}} {\bf I}_{M} = \alpha_\text a {\bf I}_{M}\\ \label{eq:R_qa:approx}
   {\bf R}_{{\bf q}_{\text a} {\bf q}_{\text a}} &\approx \left(1 - \frac{2}{\pi} \right) {\bf I}_{M}.
\end{align}
Similarly, asymptotically we have
\begin{align}
  {\bf{R}}_{{\bf x}_{\text R} {\bf x}_{\text R}}  \approx \text{diag} \left( {\bf{R}}_{{\bf x}_{\text R} {\bf x}_{\text R}} \right)
  \approx {\hat\alpha}_\text d {\bf I}_{M},
\end{align}
where
\begin{align}
&{\hat\alpha}_\text d = M \left(\alpha_\text a^2 + 1 - \frac{2}{\pi}\right) \sum\limits_{k=1}^K \sigma_{\text{SR},k}^2 \sigma_{\text{RD},k}^2 \\ \notag
&+M  \alpha_\text a^2 \sum\limits_{k=1}^K \sigma_{\text{SR},k}^2 \sigma_{\text{RD},k}^2 \left(M p_{\text S,k} \sigma_{\text{SR},k}^2 + \sum\limits_{i=1}^K p_{\text S,i} \beta_{\text{SR},i}\right).
  \end{align}
Note that the proof of calculating the approximate ${\bf{ R}}_{{\bf x}_{\text R} {\bf x}_{\text R}}$ can be found in the Appendix \ref{app:theorm:Rk_tilde}.

As a result, the matrices ${\bf{ A}}_\text d$ and ${\bf R}_{{\bf q}_{\text d} {\bf q}_{\text d}}$ can be approximated by
\begin{align}\label{eq:Ad:approx}
{\bf{ A}}_\text d &\approx \sqrt{\frac{2}{\pi{\hat\alpha}_\text d}} {\bf I}_{M}= \alpha_\text d {\bf I}_{M}\\ \label{eq:R_qd:approx}
{\bf R}_{{\bf q}_{\text d} {\bf q}_{\text d}} &\approx \left(1 - \frac{2}{\pi} \right) {\bf I}_{M}.
\end{align}
\subsection{Approximate Rate Analysis}
In this section, we derive a simpler closed-form approximation for the achievable rate. Substituting \eqref{eq:Aa:approx}, \eqref{eq:R_qa:approx}, \eqref{eq:Ad:approx}, and \eqref{eq:R_qd:approx} into \eqref{eq:rate_exact}, the exact achievable rate $R_k$ can be approximated by
\begin{align}\label{eq:rate_approx}
   &{\tilde R}_k= \\
   &\frac{\tau_\text c - 2 \tau_\text p}{2\tau_\text c} \log_2\left(1 + \frac{{\tilde A}_k}{{\tilde B}_k + {\tilde C}_k + {\tilde D}_k + {\tilde E}_k + {\tilde F}_k + {\tilde G}_k} \right), \notag
\end{align}
where
\begin{align}
{\tilde A}_k &= p_{{\text S},k} |{\tt E}\left\{ {\bf g}_{\text{RD},k}^T {\bf W} {\bf g}_{\text{SR},k}\right\} |^2, \\
{\tilde B}_k &= p_{{\text S},k} \text{Var}\left( {\bf g}_{\text{RD},k}^T {\bf W} {\bf g}_{\text{SR},k} \right), \\
{\tilde C}_k &= \sum\limits_{i\neq k} p_{{\text S},i} {\tt E} \left\{ |{\bf g}_{\text{RD},k}^T {\bf W} {\bf g}_{\text{SR},i}|^2 \right\},\\
{\tilde D}_k &= {\tt E}\left\{ ||{\bf g}_{\text{RD},k}^T {\bf W} ||^2\right\}, \\
{\tilde E}_k &= \left(1 - \frac{2}{\pi}\right)\frac{1}{\alpha_\text a^2} {\tt E}\left\{ ||{\bf g}_{\text{RD},k}^T {\bf W} ||^2\right\}, \\
{\tilde F}_k &= \left(1 - \frac{2}{\pi}\right)\frac{1}{\alpha_\text a^2 \alpha_\text d^2} {\tt E}\left\{||{\bf g}_{\text{RD},k}||^2\right\},\\
{\tilde G}_k &= \frac{1}{\gamma^2 \alpha_\text a^2 \alpha_\text d^2}.
\end{align}

With this expression, we can compute ${\tilde R}_k$ by using random matrix theory and present a closed-form approximate rate for the \emph{k}-th destination, as formalized in the following theorem.
\begin{theorem}\label{theorm:Rk_tilde}
  With one-bit ADCs and DACs at the relay, the approximate achievable rate of the \emph{k}-th destination is given by \eqref{eq:rate_approx}, where
\begin{align}
&{\tilde A}_k = p_{\text S,k} M^4 \sigma_{\text{SR},k}^4 \sigma_{\text{RD},k}^4,\\
    &{\tilde B}_k = p_{\text S,k} M^2 \left( M \sigma_{\text{SR},k}^4 \sigma_{\text{RD},k}^2 \beta_{\text{RD},k} + \beta_{\text{SR},k} t_k \right),\\
    &{\tilde C}_k = M^2 \sum\limits_{i\neq k}  p_{\text S,i} \left( M \sigma_{\text{SR},i}^4 \sigma_{\text{RD},i}^2 \beta_{\text{RD},k} + \beta_{\text{SR},i} t_k \right),\\
    &{\tilde D}_k = M^2 t_k ,\\
    &{\tilde E}_k = \left(\frac{\pi}{2} - 1\right)  \left(1 + \sum\limits_{k=1}^K p_{\text S,k} \beta_{\text{SR},k} \right) M^2 t_k, \\
  &{\tilde F}_k = \beta_{\text{RD},k}\left(\frac{\pi}{2} - 1\right)M^3 \sum\limits_{k=1}^K p_{\text S,k} \sigma_{\text{SR},k}^4 \sigma_{\text{RD},k}^2\\ \notag
 & + \beta_{\text{RD},k}\frac{M^2\pi}{2} \left(\frac{\pi}{2} - 1\right) \left(1 + \sum\limits_{k=1}^K p_{\text S,k} \beta_{\text{SR},k} \right) \sum\limits_{k=1}^K \sigma_{\text{SR},k}^2 \sigma_{\text{RD},k}^2\\
   & {\tilde G}_k = \frac{M^3\pi }{2 p_\text R} \sum\limits_{k=1}^K p_{\text S,k} \sigma_{\text{SR},k}^4 \sigma_{\text{RD},k}^2 \\ \notag
  &+ \frac{M^2 \pi^2 }{4p_\text R} \left(1 + \sum\limits_{k=1}^K p_{\text S,k} \beta_{\text{SR},k} \right) \sum\limits_{k=1}^K \sigma_{\text{SR},k}^2 \sigma_{\text{RD},k}^2,
    \end{align}
  with $t_k = M \sigma_{\text{RD},k}^4 \sigma_{\text{SR},k}^2 + \beta_{\text{RD},k} \sum\limits_{n=1}^K \sigma_{\text{SR},n}^2 \sigma_{\text{RD},n}^2$.

\end{theorem}
\proof See Appendix \ref{app:theorm:Rk_tilde}. \endproof

From Theorem \ref{theorm:Rk_tilde}, we can more readily see the impact of key parameters on the achievable rate. For instance, ${\tilde R}_k$ decreases with the number of user pairs $K$. This is expected since a higher number of users increases the amount of inter-user interference. In addition, ${\tilde R}_k$ is an increasing function of $M$, which reveals that increasing the number of relay's antennas always boosts the system performance. As $p_{\text S,k}$ approaches infinity, ${\tilde R}_k$ converges to a constant that is independent of $p_{\text S,k}$. In this case, the system becomes interference-limited.

To quantify the impact of the double quantization on system performance, in the following corollaries we compare the achievable rate with several different ADC and DAC configurations.

\begin{corollary}\label{coro:perfect:ADC}
  With perfect ADCs and one-bit DACs, the achievable rate of the \emph{k}-th destination can be expressed as \eqref{eq:rate:pA} (shown on the top of the next page),
  \begin{figure*}
  \begin{align}\label{eq:rate:pA}
    R_k^\text{pA} =
    \frac{\tau_\text c - 2 \tau_\text p}{2\tau_\text c} \log_2\left(1 + \frac{{\hat A}_k}{{\hat B}_k + {\hat C}_k + {\hat D}_k + \left(\frac{\pi}{2} - 1\right)M\beta_{\text{RD},k}{\tilde\alpha}_\text d + \frac{\pi{\tilde\alpha}_\text d}{2\gamma^2}} \right),
  \end{align}
  \hrule
  \end{figure*}
  where
  \begin{align}
  &{\tilde\alpha}_\text d =  M \sum\limits_{k=1}^K {\hat\sigma}_{\text{SR},k}^2 {\hat\sigma}_{\text{RD},k}^2\\ \notag
  &+M\sum\limits_{k=1}^K {\hat\sigma}_{\text{SR},k}^2 {\hat\sigma}_{\text{RD},k}^2 \left(M p_{\text S,k} {\hat\sigma}_{\text{SR},k}^2 + \sum\limits_{i=1}^K p_{\text S,i} \beta_{\text{SR},i}\right) ,
  \end{align}
   with ${\hat\sigma}_{\text{SR},k}^2 = \frac{K\beta_{\text{SR},k}^2p_\text p}{K\beta_{\text{SR},k}p_\text p + 1}$ and ${\hat\sigma}_{\text{RD},k}^2 = \frac{K\beta_{\text{RD},k}^2p_\text p}{K\beta_{\text{RD},k}p_\text p + 1}$; ${\hat A}_k$, ${\hat B}_k$, ${\hat C}_k$, ${\hat D}_k$ can be obtained by replacing ${\sigma}_{\text{SR},k}^2$ and ${\sigma}_{\text{RD},k}^2$ with ${\hat\sigma}_{\text{SR},k}^2$ and ${\hat\sigma}_{\text{RD},k}^2$ in ${\tilde A}_k$, ${\tilde B}_k$, ${\tilde C}_k$, ${\tilde D}_k$, respectively.
\end{corollary}

\begin{corollary}
  With perfect DACs and one-bit ADCs, the achievable rate of the \emph{k}-th destination can be expressed as
  \begin{align}\label{coro:perfect:DAC}
    R_k^\text{pD} = \frac{\tau_\text c - 2 \tau_\text p}{2\tau_\text c} \log_2\left(1 + \frac{{\tilde A}_k}{{\tilde B}_k + {\tilde C}_k + {\tilde D}_k + {\tilde E}_k + \frac{2}{\pi}{\tilde G}_k} \right).
  \end{align}
\end{corollary}
\begin{corollary}\label{coro:perfect:ADC:DAC}
With perfect ADCs and DACs, the achievable rate of the \emph{k}-th destination can be expressed as
\begin{align}\label{eq:rate:perfect}
 R_k^\text{p} = \frac{\tau_\text c - 2 \tau_\text p}{2\tau_\text c} \log_2\left(1 + \frac{{\hat A}_k}{{\hat B}_k + {\hat C}_k + {\hat D}_k + \frac{{\tilde\alpha}_\text d}{\gamma^2}} \right).
\end{align}
\end{corollary}

Corollaries \ref{coro:perfect:ADC}-\ref{coro:perfect:ADC:DAC} together with Theorem \ref{theorm:Rk_tilde} provide four cases with different ADC/DAC configurations at the relay: 1) Case I: perfect ADCs and DACs; 2) Case II: perfect ADCs and one-bit DACs; 3) Case III: one-bit ADCs and perfect DACs; 4) Case IV: one-bit ADCs and DACs. The relative performance of these four configurations is described below.

\begin{proposition}\label{prop:rate:compare}
As the number of relay antennas becomes very large, we have
\begin{align}
R_k^\text p > R_k^\text{pA} > R_k^\text{pD} > {\tilde R}_k.
\end{align}
\end{proposition}
\proof See Appendix \ref{app:prop:rate:compare}. \endproof

Proposition \ref{prop:rate:compare} indicates that the rate of the system with perfect ADCs and one-bit DACs is higher than that of one-bit ADCs and perfect DACs system. This is because one-bit ADCs cause both channel estimation errors and rate degradation, while one-bit DACs only lead to a rate reduction. For what follows, we define the three rate ratios
\begin{align}
  \left[\delta_1, \delta_2, \delta_3\right] = \left[\frac{R_k^\text{pA}}{R_k^\text p}, \frac{R_k^\text{pD}}{R_k^\text{p}}, \frac{{\tilde R}_k}{R_k^\text{p}}\right].
\end{align}

We will compare these ratios for low SNR situations where massive MIMO systems are likely to operate. Here, we consider two cases: a) the transmit power of each source scales as $p_\text S = E_\text S/M$ (where we define $p_\text S = p_{\text S,k}$ for $k = 1,\ldots,K$) with fixed $E_\text S$, while $p_\text p$ and $p_\text R$ are fixed. This case focuses on the potential power savings of the sources; b) the transmit power of the relay scales as $p_\text R = E_\text R/M$ with fixed $E_\text R$, while $p_\text S$ and $p_\text R$ are fixed. This case focuses on the potential power savings of the relay.

\begin{proposition}\label{prop:ratio_ps0}
 With $p_\text S = E_\text S/M$, and $E_\text u$, $p_\text p$, $p_\text R$ fixed, we have
 \begin{align}
 {\tilde R}_k &\rightarrow \frac{\tau_\text c - 2 \tau_\text p}{2\tau_\text c}\log_2\left(1 + \frac{2}{\pi}E_\text S \sigma_{\text{SR},k}^2 \right)\\
 R_k^\text p &\rightarrow \frac{\tau_\text c - 2 \tau_\text p}{2\tau_\text c}\log_2\left(1 + E_\text S {\hat\sigma}_{\text{SR},k}^2\right),
 \end{align}
as $M \rightarrow \infty$. In addition, if $E_\text S \rightarrow 0$, the rate ratios are given by
\begin{align}
  \left[\delta_1, \delta_2, \delta_3\right] = \left[1, 4/\pi^2, 4/\pi^2 \right].
\end{align}
\end{proposition}

\begin{proposition}\label{prop:ratio_pr0}
 With $p_\text R = E_\text R/M$, and $E_\text R$, $p_\text p$, $p_\text S$ fixed, we have
 \begin{align}
 {\tilde R}_k &\rightarrow \frac{\tau_\text c - 2 \tau_\text p}{2\tau_\text c}\log_2\left(1 + \frac{2 E_\text R}{\pi} \frac{ \sigma_{\text{SR},k}^4  \sigma_{\text{RD},k}^4}{ \sum\limits_{k=1}^K\sigma_{\text{SR},k}^4  \sigma_{\text{RD},k}^2} \right)\\
 R_k^\text p &\rightarrow \frac{\tau_\text c - 2 \tau_\text p}{2\tau_\text c}\log_2\left(1 +  E_\text R \frac{ \hat\sigma_{\text{SR},k}^4  \hat\sigma_{\text{RD},k}^4}{ \sum\limits_{k=1}^K \hat\sigma_{\text{SR},k}^4  \hat\sigma_{\text{RD},k}^2} \right),
  \end{align}
  as $M \rightarrow \infty$. In addition, if $E_\text R \rightarrow 0$, the rate ratios are given by
\begin{align}
 \left[\delta_1, \delta_2, \delta_3\right] = \left[2/\pi, 2/\pi, 4/\pi^2 \right].
\end{align}
\end{proposition}

From Propositions \ref{prop:ratio_ps0} and \ref{prop:ratio_pr0}, we can see that the system with one-bit ADCs and DACs has the same power scaling laws as the perfect hardware case, which is an encouraging result. In addition, for both Propositions \ref{prop:ratio_ps0} and \ref{prop:ratio_pr0}, $\delta_3 = 4/\pi^2$, revealing that for the double-quantized system, the rate ratio is $4/\pi^2$ times less than the perfect ADC/DAC case, for low transmit power at the sources or low transmit power at the relay. This result is the same as that in the system which is only quantized once \cite{A.Mezghani2}. Interestingly, focusing on the values of $\delta_1$ and $\delta_2$, we observe that the process to achieve the final scaling of $4/\pi^2$ is quite different. For low $p_{\text S,k}$ case, the value $4/\pi^2$ only results from $\delta_2 = 4/\pi^2$, implying that the rate loss is only caused by the one-bit ADCs. In contrast, for the low $p_\text R$ case, the value $4/\pi^2$ is generated by $\delta_1 = 2/\pi$ and $\delta_2 = 2/\pi$, indicating that the rate degradation comes from both one-bit ADCs and one-bit DACs.
\section{Power Allocation}\label{sec:power_allocation}
In this section, we formulate a power allocation problem maximizing the sum rate of the system for a given total power budget $P_\text T$, i.e., $\sum\limits_{k=1}^K p_{\text S,k} + p_\text R \leq P_\text{T}$.
\subsection{Problem Formulation}
Defining ${\bf p}_\text S = \left[p_{\text S,1}, \ldots, p_{\text S,K} \right]^T$, the problem is expressed as

\begin{align}\label{problem1}
{\cal P}_1:  \mathop{\text{maximize}}\limits_{{\bf p}_\text S, p_r} \quad &\frac{\tau_c - 2 \tau_p}{2 \tau_c} \sum\limits_{k = 1}^{K} \log_2  \left( 1 + \gamma_k \right) \\ \label{eq:gamma:equality}
  {\text{subject to}} \quad  &\gamma_k = \frac{p_{\text S,k}}{\xi_k}, k = 1, \ldots, K\\
  &\sum\limits_{i = 1}^K p_{\text S,i}  + p_r \leq P_\text T\\
  &{\bf p}_\text S \geq {\bf 0}, p_r \geq 0,
\end{align}
where
\begin{align}
  \xi_k &= \sum\limits_{i=1}^K p_{\text S,i} a_{k,i} + p_\text R^{-1} \left( \sum\limits_{i=1}^K b_{k,i} p_{\text S,i} + c_k \right) + d_k\\
  c_k &= \frac{\pi^2 }{4 M^2\sigma_{\text{SR},k}^4 \sigma_{\text{RD},k}^4} \sum\limits_{n=1}^K \sigma_{\text{SR},n}^2 \sigma_{\text{RD},n}^2\\
  d_k &=  \frac{\pi}{2 M\sigma_{\text{SR},k}^2} + \frac{\pi^2 \beta_{\text{RD},k}}{4 M^2 \sigma_{\text{SR},k}^4 \sigma_{\text{RD},k}^4}
\sum\limits_{n = 1}^K \sigma_{\text{SR},n}^2 \sigma_{\text{RD},n}^2,
\end{align}
and $a_{k,i}$ and $b_{k,i}$ are respectively given by \eqref{eq:a_k} and \eqref{eq:b_k}, shown on the top of the next page.
\begin{figure*}
\begin{align}\label{eq:a_k}
a_{k,i} =
\begin{cases}
\frac{\pi}{2 M\sigma_{\text{SR},k}^2 \sigma_{\text{RD},k}^2} \left( \sigma_{\text{SR},k}^2 \beta_{\text{RD},k}
+ \sigma_{\text{RD},k}^2 \beta_{\text{SR},k} \right) + \frac{\pi^2 \beta_{\text{SR},k} \beta_{\text{RD},k} }{4 M^2 \sigma_{\text{SR},k}^4 \sigma_{\text{RD},k}^4}  \sum\limits_{n=1}^K \sigma_{\text{SR},n}^2 \sigma_{\text{RD},n}^2, & i = k  \\
\frac{1}{M\sigma_{\text{SR},k}^4 \sigma_{\text{RD},k}^4}  \left(\sigma_{\text{SR},k}^2 \sigma_{\text{RD},k}^4 \beta_{\text{SR},i} + \frac{\pi}{2} \beta_{\text{RD},k} \sigma_{\text{SR},i}^4 \sigma_{\text{RD},i}^2  + \left(\frac{\pi}{2} - 1\right) \beta_{\text{SR},i} \sigma_{\text{SR},i}^2 \sigma_{\text{RD},i}^4 \right) \\ + \frac{\beta_{\text{SR},i}}{M^2 \sigma_{\text{SR},k}^4 \sigma_{\text{RD},k}^4} \left( \left( \frac{\pi^2}{4} - \frac{\pi}{2} + 1 \right) \beta_{\text{RD},k} + \left(\frac{\pi}{2} - 1\right) \beta_{\text{RD},i} \right) \sum\limits_{n = 1}^K \sigma_{\text{SR},n}^2 \sigma_{\text{RD},n}^2 , & i \neq k,
\end{cases}
\end{align}
\hrule
\end{figure*}
\begin{figure*}
\begin{align}\label{eq:b_k}
b_{k,i} =
\begin{cases}
\frac{\pi }{2} \left(\frac{1}{M \sigma_{\text{RD},k}^2 } + \frac{\pi \beta_{\text{SR},k}}{2 M^2\sigma_{\text{SR},k}^4 \sigma_{\text{RD},k}^4} \sum\limits_{n=1}^K \sigma_{\text{SR},n}^2 \sigma_{\text{RD},n}^2 \right), &i = k\\
\frac{\pi}{2 M^2 \sigma_{\text{SR},k}^4 \sigma_{\text{RD},k}^4} \left(M \sigma_{\text{SR},i}^4 \sigma_{\text{RD},i}^2 + \frac{\pi}{2} \beta_{\text{SR},i} \sum\limits_{n=1}^K \sigma_{\text{SR},n}^2 \sigma_{\text{RD},n}^2 \right), &i\neq k,
\end{cases}
\end{align}
\hrule
\end{figure*}

Since $\log\left(\cdot\right)$ is an increasing function, problem ${\cal P}_1$ can be reformulated as
\begin{align}\label{problem2}
{\cal P}_2:  \mathop{\text{minimize}}\limits_{{\bf p}_\text S, p_r} \quad & \prod_{k = 1}^{K} \left( 1 + \gamma_k \right)^{-1} \\ \label{eq:gamma:inequality}
  {\text{subject to}} \quad &\gamma_k \leq \frac{p_{\text S,k}}{\xi_k} , k = 1, \ldots, K \\
   &\sum\limits_{i = 1}^K p_{\text S,i}  + p_r \leq P_\text T\\
  &{\bf p}_\text S \geq {\bf 0}, p_r \geq 0,
\end{align}
which can be identified as a complementary geometric program (CGP) \cite{M.Avriel}. Note that the equality constraints \eqref{eq:gamma:equality} of ${\cal P}_1$ have been replaced with inequality constraints \eqref{eq:gamma:inequality}. Since the objective function of ${\cal P}_2$ decreases with $\gamma_k$, we can guarantee that the inequality constraints \eqref{eq:gamma:inequality} must be active at any optimal solution of ${\cal P}_2$, which means that problem ${\cal P}_2$ is equivalent to ${\cal P}_1$.

\subsection{Successive Approximation Algorithm}
CGP problems are in general nonconvex. Fortunately, we can first approximate the CGP by solving a sequence of GP problems. Then, each GP can be solved very efficiently with standard convex optimization tools such as CVX. The key idea is to use a monomial function $\omega_k\gamma_k^{\mu_k}$ to approximate $1 + \gamma_k$ near an arbitrary point ${\hat \gamma}_k > 0$. To make the approximation accurate, we need to ensure that
\begin{align}
\begin{cases}
  1 + {\hat\gamma}_k = \omega_k{\hat\gamma_k}^{\mu_k}\\
  {\mu_k}\omega_k{\hat\gamma}_k^{\mu_k - 1} = 1.
\end{cases}
\end{align}
These results will hold if the parameters $\omega_k$ and $\mu_k$ are chosen as $\omega_k = {\hat\gamma}_k^{-\mu_k} \left(1 + {\hat\gamma}_k\right)$ and $\mu_k = \frac{{\hat\gamma}_k}{1+{\hat\gamma}_k}$. At each iteration, the GP is obtained by replacing the posynomial objective function with its best local monomial approximation near the solution obtained at the previous iteration. The following algorithm shows the steps to solving ${\cal P}_2$.

\begin{algorithm}[ht!]
  \caption{Successive approximation algorithm for ${\cal{P}}_2$}\label{algorithm}

    1) {\it Initialization}. Define a tolerance $\epsilon$ and parameter $\theta$. Set $j = 1$, and set the initial value of ${\hat\gamma}_k$ according to the signal-to-interference-plus-noise ratio (SINR) in Theorem \ref{theorm:Rk_tilde} with $p_{\text S,k} = \frac{P_\text T}{2K}$ and $p_r = \frac{P_\text T}{2}$.

    2) {\it Iteration $j$}. Compute $\mu_k = \frac{{\hat\gamma}_k}{1+{\hat\gamma}_k}$. Then, solve the following GP problem ${\cal{P}}_3$:
\begin{align}\label{problem4}
{\cal P}_3:  \mathop{\text{minimize}}\limits_{{\bf p}_\text S, p_r} \quad & \prod_{k = 1}^{K} \gamma_k^{-\mu_k} \\
  {\text{subject to}} \quad &\theta^{-1}{\hat\gamma}_k \leq {\gamma}_k \leq \theta {\hat\gamma}_k, k = 1, \ldots, K\\
  &\gamma_k p_{\text S,k}^{-1} \xi_k \leq 1, k = 1, \ldots, K \\
  &\sum\limits_{i = 1}^K p_{\text S,i}  + p_r \leq P_\text T\\
  &{\bf p}_\text S \geq {\bf 0}, p_r \geq 0.
\end{align}

Denote the optimal solutions by ${\gamma}_{k}^{(j)}$, for $k = 1,\ldots, K$.

    3) {\it Stopping criterion}. If $\max_k |{\gamma}_k^{(j)} - {\hat\gamma}_{k}| < \epsilon$, stop; otherwise, go to step 4).

    4) {\it Update initial values}. Set ${\hat\gamma}_k = {\gamma}_k^{(j)}$, and $j = j + 1$. Go to step 2).
\end{algorithm}

We have neglected $\omega_k$ in the objective function of ${\cal P}_3$ since they are constants and do not affect the problem solution. Also, some trust region constraints are added, i.e., $\theta^{-1}{\hat\gamma}_k \leq {\gamma}_k \leq \theta {\hat\gamma}_k$, which limits how much the variables are allowed to differ from the current guess ${\hat\gamma}_k$. The parameter $\theta > 1$ controls the desired accuracy. More precisely, when $\theta$ is close to 1, it provides good accuracy for the momomial approximation but with slower convergence speed, and vice versa if $\theta$ is large. As discussed in \cite{S.Boyd2}, $\theta = 1.1$ offers a good tradeoff between accuracy and convergence speed.
\section{Numerical Results}\label{sec:numerical_results}
In this section, we present numerical results to validate previous analytical results and demonstrate the benefits of the power allocation algorithm.

\subsection{Impact of the input pilot matrix}
In this section, we evaluate the channel estimation accuracy of the identity and Hadamard pilot matrices. We choose $K = 4$, and the large scale fading coefficients ${\bf \beta}_\text{SR} = [0.6, 0.3, 0.1, 0.9]$.

\begin{figure}[!ht]
    \centering
    \includegraphics[scale=0.45]{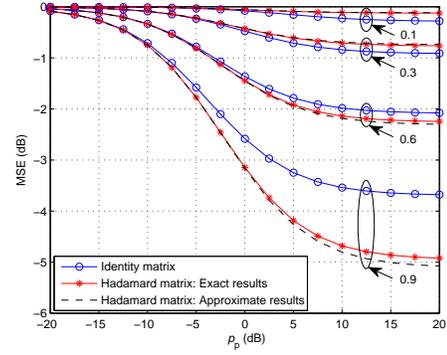}
    \caption{MSE versus $p_\text p$ for $K = 4$ and $M = 128$.}\label{fig:MSE_pp}
  \end{figure}

Fig. \ref{fig:MSE_pp} illustrates the MSE of each channel from the sources to the relay versus the transmit power of each pilot symbol. For $\beta_{\text{SR},k} = \left\{0.1, 0.3\right\}$ which are less than the average large scale fading value of $0.475$, the identity matrix pilot outperforms the Hadamard matrix, in agreement with Proposition \ref{theor:pilot_matrix}. In addition, observing the curves associated with the Hadamard matrix, we can see that the approximate results nearly overlap with the exact results in the low $p_\text p$ regime, indicating the validity of our theoretical analysis. However, if $p_\text p$ increases, the gap between the approximate and exact results grows.

\subsection{Validation of analytical results}
In this section, we validate the theoretical derivations. For simplicity, we set the large-scale fading coefficients as $\beta_{\text{SR},k} = \beta_{\text{RD},k} = 1$ and adopt an equal power allocation strategy, i.e., $p_{\text S,k} = p_\text S$.

\begin{figure}[!ht]
    \centering
    \includegraphics[scale=0.45]{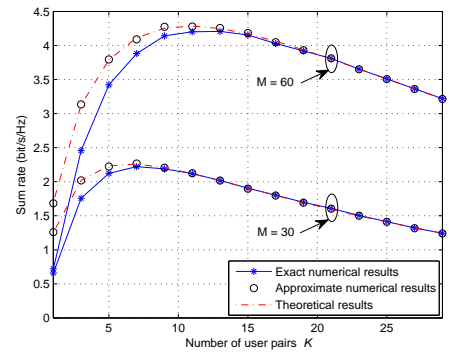}
    \caption{Sum rate versus the the number of user pairs $K$ for $p_\text S = 10$ dB, $p_\text R = 10$ dB, and $p_p = 10$ dB.}\label{fig:rate_num_user_ps10}
  \end{figure}

Fig. \ref{fig:rate_num_user_ps10} shows the sum rate versus the number of user pairs $K$. The curves associated with ``Exact numerical results'' and ``Approximate numerical results'' are respectively generated by Monte-Carlo simulations according to \eqref{eq:rate_exact} and \eqref{eq:rate_approx} by averaging over $10^3$ independent channel realizations, and the ``Theoretical results'' curves are obtained based on Theorem \ref{theorm:Rk_tilde}. As can be seen, there exists a gap between ``Exact numerical results'' (where the matrices ${\bf R}_{{\bf q}_\text a {\bf q}_\text a}$ and ${\bf R}_{{\bf q}_\text d {\bf q}_\text d}$ are not diagonal, which means that the quantization noise is correlated) and ``Approximate numerical results'' (where the matrices ${\bf R}_{{\bf q}_\text a {\bf q}_\text a}$ and ${\bf R}_{{\bf q}_\text d {\bf q}_\text d}$ are approximated by identity matrices) when the number of user pairs is small, while the gap narrows and finally disappears as $K$ becomes large. The reason is that the correlation effect is stronger with smaller $K$ and weaker with larger $K$. In this example, our approximate model is very accurate when the number of user pairs is greater than $15$, which is a reasonable number for this size of array. In addition, we observe that the ``Approximate numerical results'' curve overlaps with that for the ``Theoretical results'', which verifies our analytical derivations in Theorem \ref{theorm:Rk_tilde}.

\begin{figure}[ht]
  \centering
  \subfigure[$K = 10$]{\label{fig:rate_num_antenna_K10_ps10_pr10}\includegraphics[width=0.37\textwidth]{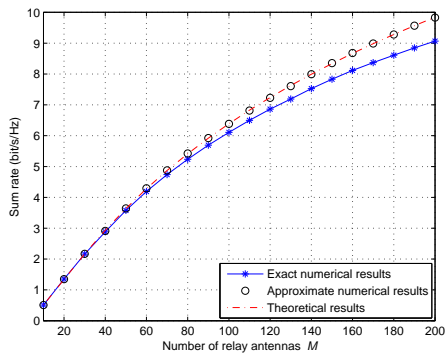}}
  \hspace{0.2in}
  \subfigure[$K = M/10$]{\label{fig:rate_num_antenna_MdK10_ps10_pr10}\includegraphics[width=0.37\textwidth]{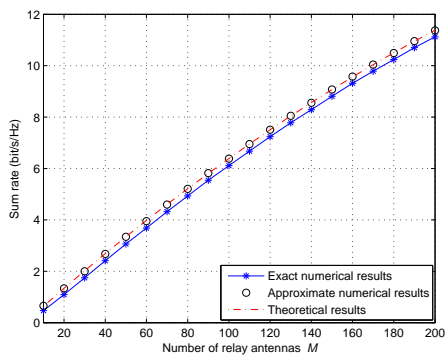}}
  \caption{Sum rate versus the number of relay antennas $M$ for $p_\text S = 10$ dB, $p_\text R = 10$ dB, and $p_p = 10$ dB.}\label{fig:rate_num_antenna_ps10_pr10}
\end{figure}

Fig. \ref{fig:rate_num_antenna_ps10_pr10} shows the sum rate versus the number of relay antennas. From Fig. \ref{fig:rate_num_antenna_K10_ps10_pr10}, we can see that when $K = 10$, the gap between the exact and approximate numerical results increases with the number of relay antennas. This suggests that for large antenna arrays, the correlation of the quantization noise becomes important and cannot be neglected. However, we are interested in the typical massive MIMO setup where the ratio between the number of relay antennas and the users is on the order of about $M/K = 10$, and thus we plot Fig. \ref{fig:rate_num_antenna_MdK10_ps10_pr10}. In this figure, we can see the gap slightly narrows (from 0.2791 bit/s/Hz at $M = 80$ to 0.2505 bit/s/Hz at $M = 200$) as the number of relay antennas increases, which indicates that our approximate model is accurate for massive MIMO scenarios.

\begin{figure}[ht]
  \centering
  \subfigure[Required $p_\text S$]{\label{fig:required_ps}\includegraphics[width=0.37\textwidth]{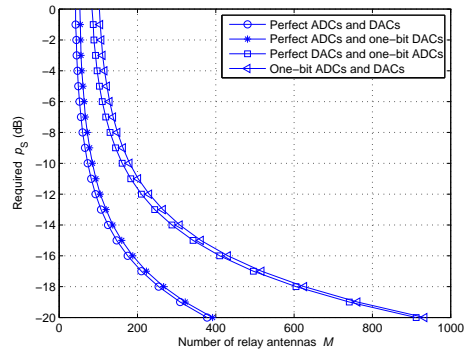}}
  \hspace{0.2in}
  \subfigure[Rate ratio: $p_\text S = -50$ dB]{\label{fig:rate_ratio_ps0}\includegraphics[width=0.37\textwidth]{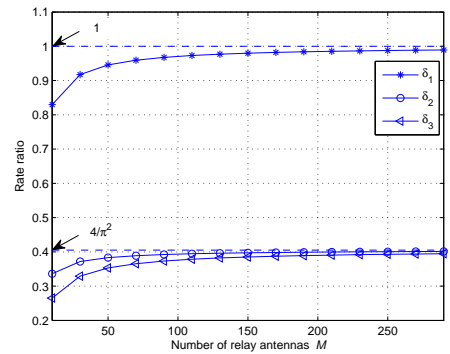}}
  \caption{Required $p_\text S$ and rate ratio versus the number of relay antennas $M$ for $K = 5$, $p_p = 10$ dB, and $p_\text R = 10$ dB.}\label{fig:rate_ps0}
\end{figure}

Fig. \ref{fig:required_ps} shows the transmit power $p_\text S$ of each source required to maintain a given sum rate of $5$ bit/s/Hz. We can see that when the number of relay antennas increases, the required $p_\text S$ is significantly reduced. Furthermore, if the number of relay antennas is very large, the required $p_\text S$ is irrelevant to the resolution of the DACs. In other words, the sources transmit the same power in Case I and Case II, and pay the same power in Case III and Case IV. Fig. \ref{fig:rate_ratio_ps0} plots the three rate ratios versus the number of relay antennas when $p_\text S$ is very low. We observe that the three rate ratio curves converge to two nonzero limits $1$ and $4/\pi^2$, which is consistent with Proposition \ref{prop:ratio_ps0}. This property provides an efficient way to predict the sum rate with one-bit quantization according to the known sum rate of perfect ADC and/or DAC systems in low source transmit power regimes and with large-scale relay antennas.

\begin{figure}[ht]
  \centering
  \subfigure[Required $p_\text R$]{\label{fig:required_pr}\includegraphics[width=0.37\textwidth]{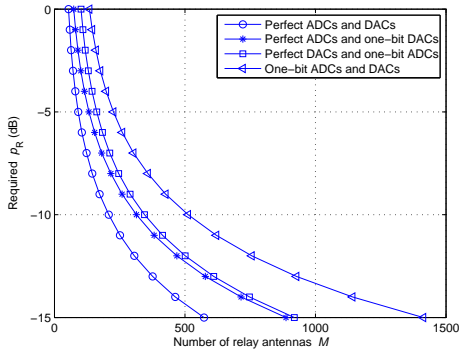}}
  \hspace{0.2in}
  \subfigure[Rate ratio: $p_\text R = -40$ dB]{\label{fig:rate_ratio_pr0}\includegraphics[width=0.37\textwidth]{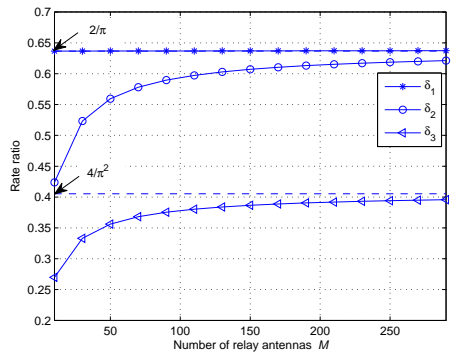}}
  \caption{Required $p_\text R$ and rate ratio versus the number of relay antennas $M$ for $K = 5$, $p_p = 10$ dB, and $p_\text S = 10$ dB.}\label{fig:rate_pr0}
\end{figure}

 Fig. \ref{fig:required_pr} shows the transmit power $p_\text R$ of the relay required to maintain a given sum rate of $5$ bit/s/Hz. As in the previous case, the required power is substantially reduced when the number of relay antennas grows, which indicates the great benefits of employing large antenna arrays. In addition, when $p_\text R$ is very small, e.g., $p_\text R = -10$ dB, the four curves show quite different results. The required number of relay antennas with one-bit ADCs and DACs is $M = 512$, which is approximately 2.5 times more than the case with perfect ADCs and DACs which requires $M = 208$ antennas. For Case II and Case III, the required number of relay antennas is almost the same, respectively $M = 314$ and $M = 345$. Fig. \ref{fig:rate_ratio_pr0} compares the three rate ratios when $p_\text R$ is very low. We can see that the three rate ratio curves converge to three nonzero limits $2/\pi$, $2/\pi$, and $4/\pi^2$, which agrees with Proposition \ref{prop:ratio_pr0}.

\subsection{Power allocation}
\begin{figure}[ht]
\centering
 \includegraphics[scale=0.45]{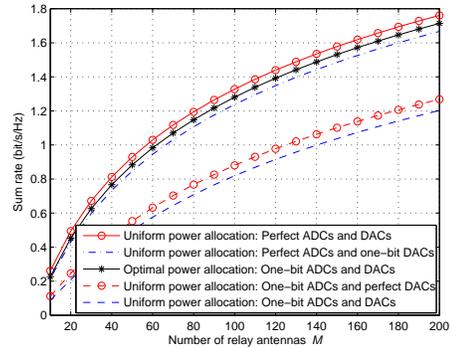}
  \caption{Sum rate versus the number of relay antennas $M$ for $K = 5$, $p_p = 10$ dB, and $P_\text T = 10$ dB.}\label{fig:power_allocation}
\end{figure}

Fig. \ref{fig:power_allocation} illustrates the impact of the optimal power allocation scheme on the sum rate when all users experience different large-scale fading. The large-scale fading coefficients are arbitrarily generated by $\beta_{\text{SR},k} = z_k\left(r_{\text{SR},k}/r_0 \right)^\kappa$ and $\beta_{\text{RD},k} = z_k\left(r_{\text{RD},k}/r_0 \right)^\kappa$, where $z_k$ is a log-normal random variable with standard deviation 8 dB, $r_{\text{SR},k}$ and $r_{\text{RD},k}$ respectively represent the distances from the sources and destinations to the relay, $\kappa = 3.8$ is the path loss exponent, and $r_0$ denotes the guard interval which specifies the nearest distance between the users and the relay. The relay is located at the center of a cell with a radius of $1000$ meters and $r_0 = 100$ meters. We choose $\beta_\text{SR} = \left[0.2688, 0.0368, 0.00025, 0.1398, 0.0047 \right]$, and $\beta_\text{RD} = \left[0.0003, 0.00025, 0.0050, 0.0794, 0.0001 \right]$. As a benchmark scheme for comparison, we also plot the sum rate with uniform power allocation, i.e., $p_\text S = \frac{P_\text T}{2K}$ and $p_\text R = \frac{P_\text T}{2}$. For uniform power allocation, we can see that the rate of Case I is the highest, Case IV is the lowest, while Case II outperforms Case III. These results are in agreement with Proposition \ref{prop:rate:compare}. In addition, we observe that the optimal power allocation strategy significantly boosts the sum rate. Although the rate achieved by the optimal power allocation with one-bit ADCs and DACs is inferior to the case of perfect ADCs and DACs with uniform power allocation, it outperforms the other three one-bit ADC/DAC configurations. This demonstrates the great importance of power allocation in quantized systems.

\section{Conclusions}\label{sec:conclusions}
We have analyzed the achievable rate of a multipair half-duplex massive antenna relaying system assuming that one-bit ADCs and DACs are deployed at the relay. An approximate closed-form expression for the achievable rate was derived, based on which the impact of key system parameters was characterized. It was shown that  the sum rate with one-bit ADCs and DACs is $4/\pi^2$ times less than that achieved by an unquantized system in the low power regime. Despite the rate loss due to the use of one-bit ADCs and DACs, employing massive antenna arrays still enables significant power savings; i.e., the transmit power of each source or the relay can be reduced proportional to $1/M$ to maintain a constant rate, as in the unquantized case. Finally, we show that a good power allocation strategy can substantially compensate for the rate loss caused by the coarse quantization.

\appendices
\section{Proof of Theorem \ref{theorm:Rk_tilde}}\label{app:theorm:Rk_tilde}
The end-to-end SINR given in \eqref{eq:rate_approx} consists of six expectation terms: 1) desired signal power ${\tilde A}_k$; 2) estimation error ${\tilde B}_k$; 3) interpair interference ${\tilde C}_k$; 4) noise at the relay ${\tilde D}_k$; 5) quantization noise of ADCs ${\tilde E}_k$; 6) quantization noise of DACs ${\tilde F}_k$. Besides these terms, we also need to calculate an approximation of ${\bf{R}}_{{\bf x}_{\text R} {\bf x}_{\text R}}$. In the following, we compute them one by one.

1) Approximate ${\bf{R}}_{{\bf x}_{\text R} {\bf x}_{\text R}}$:
\begin{align}\label{eq:R_tilde_xR}
  &{\bf{ R}}_{{\bf x}_{\text R} {\bf x}_{\text R}} = {\tt E}\left\{ {\bf{\hat G}}_\text{RD}^* {\bf {\hat G}}_\text{SR}^H  {\bf R}_{{\bf{\tilde y}}_{\text R} {\bf{\tilde y}}_{\text R}} {\bf{\hat G}}_\text{SR} {\bf{\hat G}}_\text{RD}^T \right\} \\ \notag
  &\approx \alpha_\text a^2 {\tt E}\left\{ {\bf{\hat G}}_\text{RD}^* {\bf{\hat G}}_\text{SR}^H {\bf G}_\text{SR} {\bf P}_\text S {\bf G}_\text{SR}^H {\bf{\hat G}}_\text{SR} {\bf{\hat G}}_\text{RD}^T \right\} \\ \notag
  &+ \left(\alpha_\text a^2 + 1 - \frac{2}{\pi}\right) {\tt E}\left\{ {\bf{\hat G}}_\text{RD}^* {\bf{\hat G}}_\text{SR}^H {\bf{\hat G}}_\text{SR} {\bf{\hat G}}_\text{RD}^T \right\}.
\end{align}
By using the fact that ${\tt E}\left\{||{\bf g}_{\text{SR},k}||^4 \right\} = M \left(M + 1\right)\beta_{\text{SR},k}^2$, we have
\begin{align}\label{eq:gamma:term1}
  & {\tt E}\left\{ {\bf{\hat G}}_\text{RD}^* {\bf{\hat G}}_\text{SR}^H {\bf G}_\text{SR} {\bf P}_\text S {\bf G}_\text{SR}^H {\bf{\hat G}}_\text{SR} {\bf{\hat G}}_\text{RD}^T \right\} \\ \notag
  &= {\tt E}\left\{ {\bf{\hat G}}_\text{RD}^* {\bf{\hat G}}_\text{SR}^H {\bf {\hat G}}_\text{SR} {\bf P}_\text S {\bf{\hat G}}_\text{SR}^H {\bf{\hat G}}_\text{SR} {\bf{\hat G}}_\text{RD}^T \right\} \\ \notag
  &+ {\tt E}\left\{ {\bf{\hat G}}_\text{RD}^* {\bf{\hat G}}_\text{SR}^H {\bf E}_\text{SR} {\bf P}_\text S {\bf E}_\text{SR}^H {\bf{\hat G}}_\text{SR} {\bf{\hat G}}_\text{RD}^T \right\} \\ \notag
  &= M \sum\limits_{k=1}^K \sigma_{\text{SR},k}^2 \sigma_{\text{RD},k}^2 \left(M p_{\text S,k} \sigma_{\text{SR},k}^2 + \sum\limits_{i=1}^K p_{\text S,i} \beta_{\text{SR},i} \right) {\bf I}_{M} \\ \label{eq:gamma:term2}
  & {\tt E}\left\{ {\bf{\hat G}}_\text{RD}^* {\bf{\hat G}}_\text{SR}^H {\bf{\hat G}}_\text{SR} {\bf{\hat G}}_\text{RD}^T \right\} = M\sum\limits_{k=1}^K \sigma_{\text{SR},k}^2 \sigma_{\text{RD},k}^2 {\bf I}_{M}.
\end{align}
Then, by substituting \eqref{eq:gamma:term1} and \eqref{eq:gamma:term2} into \eqref{eq:R_tilde_xR}, we directly obtain
\begin{align}\label{eq:Ad_approx}
  &{\bf{R}}_{{\bf x}_{\text R} {\bf x}_{\text R}} \approx \\ \notag
  &M \alpha_\text a^2 \sum\limits_{k=1}^K \sigma_{\text{SR},k}^2 \sigma_{\text{RD},k}^2 \left(M p_{\text S,k} \sigma_{\text{SR},k}^2 + \sum\limits_{i=1}^K p_{\text S,i} \beta_{\text{SR},i}\right){\bf I}_{M} \\
  &+ \left(\alpha_\text a^2 + 1 - \frac{2}{\pi}\right) \sum\limits_{k=1}^K \sigma_{\text{SR},k}^2 \sigma_{\text{RD},k}^2  {\bf I}_{M}. \notag
\end{align}

2) ${\tilde A}_k$: Since
\begin{align}
&{\tt E}\left\{ {\bf g}_{\text{RD},k}^T {\bf W} {\bf g}_{\text{SR},k}\right\} = {\tt E}\left\{ {\bf g}_{\text{RD},k}^T {\bf{\hat g}}_{\text{RD},k}^* {\bf{\hat g}}_{\text{SR},k}^H {\bf g}_{\text{SR},k}\right\} \\ \notag
&= M^2 \sigma_{\text{SR},k}^2 \sigma_{\text{RD},k}^2,
\end{align}
we have
\begin{align}
{\tilde A}_k = p_{\text S,k} M^4 \sigma_{\text{SR},k}^4 \sigma_{\text{RD},k}^4.
\end{align}

3) ${\tilde B}_k$:
\begin{align}
&{\tt E}\left\{ |{\bf g}_{\text{RD},k}^T {\bf{\hat G}}_\text{RD}^* {\bf{\hat G}}_\text{SR}^H {\bf g}_{\text{SR},k}|^2 \right\} = \\ \notag
& {\tt E}\left\{ \sum\limits_{m = 1}^K \sum\limits_{n=1}^K {\bf g}_{\text{RD},k}^T {\bf{\hat g}}_{\text{RD},m}^* {\bf {\hat g}}_{\text{SR},m}^H {\bf g}_{\text{SR},k} {\bf g}_{\text{SR},k}^H {\bf{\hat g}}_{\text{SR},n} {\bf{\hat g}}_{\text{RD},n}^T {\bf g}_{\text{RD},k}^* \right\},
 \end{align}
which can be decomposed into three different cases:

a) for $m = n = k$,
\begin{align}
&{\tt E}\left\{ |{\bf g}_{\text{RD},k}^T {\bf{\hat G}}_\text{RD}^* {\bf{\hat G}}_\text{SR}^H {\bf g}_{\text{SR},k}|^2 \right\}\\ \notag
&= {\tt E}\left\{ ||{\bf{\hat g}}_{\text{SR},k}||^4 ||{\bf{\hat g}}_{\text{RD},k}||^4 \right\} \\ \notag
 &+ {\tt E}\left\{ ||{\bf{\hat g}}_{\text{SR},k}||^4 |{\bf{\hat g}}_{\text{RD},k}^T {\bf e}_{\text{RD},k}^* |^2 \right\} \\ \notag
&+ {\tt E}\left\{ ||{\bf{\hat g}}_{\text{RD},k}||^4 |{\bf{\hat g}}_{\text{SR},k}^H {\bf e}_{\text{SR},k} |^2 \right\} \\ \notag
&+ {\tt E}\left\{ |{\bf{\hat g}}_{\text{SR},k}^H {\bf e}_{\text{SR},k} |^2 |{\bf{\hat g}}_{\text{RD},k}^T {\bf e}_{\text{RD},k}^* |^2 \right\}\\ \notag
&= M^2 \left(M+1\right)^2 \sigma_{\text{SR},k}^4 \sigma_{\text{RD},k}^4 \\ \notag
&+ M^2 \left(M+1\right) \sigma_{\text{SR},k}^4 \sigma_{\text{RD},k}^2 {\tilde\sigma}_{\text{RD},k}^2 \\ \notag
&+ M^2 \left(M+1\right) \sigma_{\text{RD},k}^4 \sigma_{\text{SR},k}^2 {\tilde\sigma}_{\text{SR},k}^2 \\ \notag
&+ M^2 \sigma_{\text{SR},k}^2 {\tilde\sigma}_{\text{SR},k}^2 \sigma_{\text{RD},k}^2 {\tilde\sigma}_{\text{RD},k}^2.
\end{align}

b) for $m = n \neq k$,
\begin{align}
  &{\tt E}\left\{ |{\bf g}_{\text{RD},k}^T {\bf{\hat G}}_\text{RD}^* {\bf{\hat G}}_\text{SR}^H {\bf g}_{\text{SR},k}|^2 \right\} \\ \notag
  &= M^2 \beta_{\text{SR},k} \beta_{\text{RD},k} \sum\limits_{n\neq k} \sigma_{\text{SR},n}^2 \sigma_{\text{RD},n}^2.
\end{align}

c) for $m \neq n \neq k$,
\begin{align}
{\tt E}\left\{ |{\bf g}_{\text{RD},k}^T {\bf{\hat G}}_\text{RD}^* {\bf{\hat G}}_\text{SR}^H {\bf g}_{\text{SR},k}|^2 \right\} = 0.
\end{align}

Combining a), b), and c), and by utilizing the fact of $\sigma_{\text{SR},k}^2 + {\tilde\sigma}_{\text{SR},k}^2 = \beta_{\text{SR},k}$ and $\sigma_{\text{RD},k}^2 + {\tilde\sigma}_{\text{RD},k}^2 = \beta_{\text{RD},k}$, we have
\begin{align}
  &{\tt E}\left\{ |{\bf g}_{\text{RD},k}^T {\bf{\hat G}}_\text{RD}^* {\bf{\hat G}}_\text{SR}^H {\bf g}_{\text{SR},k}|^2 \right\} \\ \notag
  &= p_{\text S,k} M^4 \sigma_{\text{SR},k}^4 \sigma_{\text{RD},k}^4 + p_{\text S,k} M^3 \sigma_{\text{SR},k}^4 \sigma_{\text{RD},k}^2 \beta_{\text{RD},k} \\ \notag
&+ p_{\text S,k} M^3 \sigma_{\text{RD},k}^4 \sigma_{\text{SR},k}^2 \beta_{\text{SR},k} \\ \notag
&+ p_{\text S,k} M^2 \beta_{\text{SR},k} \beta_{\text{RD},k} \sum\limits_{n=1}^K \sigma_{\text{SR},n}^2 \sigma_{\text{RD},n}^2.
\end{align}
Thus,
\begin{align}
  &{\tilde B}_k = p_{\text S,k} M^3 \left(\sigma_{\text{SR},k}^4 \sigma_{\text{RD},k}^2 \beta_{\text{RD},k}
+\sigma_{\text{RD},k}^4 \sigma_{\text{SR},k}^2 \beta_{\text{SR},k} \right) \\ \notag
&+  p_{\text S,k} M^2 \beta_{\text{SR},k} \beta_{\text{RD},k} \sum\limits_{n=1}^K \sigma_{\text{SR},n}^2 \sigma_{\text{RD},n}^2.
\end{align}
4) ${\tilde C}_k$:
\begin{align}
&{\tt E}\left\{ |{\bf g}_{\text{RD},k}^T {\bf{\hat G}}_\text{RD}^* {\bf {\hat G}}_\text{SR}^H {\bf g}_{\text{SR},i}|^2 \right\} =\\ \notag
& {\tt E}\left\{ \sum\limits_{m = 1}^K \sum\limits_{n=1}^K {\bf g}_{\text{RD},k}^T {\bf{\hat g}}_{\text{RD},m}^* {\bf{\hat g}}_{\text{SR},m}^H {\bf g}_{\text{SR},i} {\bf g}_{\text{SR},i}^H {\bf{\hat g}}_{\text{SR},n} {\bf{\hat g}}_{\text{RD},n}^T {\bf g}_{\text{RD},k}^* \right\},
\end{align}
which can be decomposed as six cases:

a) for $m \neq n \neq k,i$,
\begin{align}
{\tt E}\left\{|{\bf g}_{\text{RD},k}^T {\bf{\hat G}}_\text{RD}^* {\bf{\hat G}}_\text{SR}^H {\bf g}_{\text{SR},i}|^2 \right\} = 0.
\end{align}

b) for $m = n \neq k,i$,
\begin{align}
&{\tt E}\left\{ \sum\limits_{n \neq k,i} | {\bf{\hat g}}_{\text{SR},n}^H {\bf g}_{\text{SR},i} |^2 | {\bf g}_{\text{RD},k}^T {\bf{\hat g}}_{\text{RD},n}^* |^2 \right\} \\ \notag
  &= M^2 \beta_{\text{SR},i} \beta_{\text{RD},k} \sum\limits_{n \neq i,k} \sigma_{\text{SR},n}^2 \sigma_{\text{RD},n}^2.
\end{align}

c) for $m = n = k$ ($k\neq i$),
\begin{align}
  &{\tt E}\left\{ |{\bf{\hat g}}_{\text{SR},k}^H {\bf g}_{\text{SR},i}|^2 |{\bf g}_{\text{RD},k}^T {\bf{\hat g}}_{\text{RD},k}^*|^2 \right\} \\ \notag
  &= M^2 \sigma_{\text{SR},k}^2 \sigma_{\text{RD},k}^2 \beta_{\text{SR},i} \left(\left(M + 1\right)\sigma_{\text{RD},k}^2 + {\tilde\sigma}_{\text{RD},k}^2 \right).
\end{align}

d) for $m = n = i$ ($i\neq k$),
\begin{align}
  &{\tt E}\left\{ |{\bf g}_{\text{RD},k}^T {\bf{\hat g}}_{\text{RD},i}^*|^2 |{\bf{\hat g}}_{\text{SR},i}^H {\bf g}_{\text{SR},i} |^2  \right\}  \\ \notag
  &= M^2 \sigma_{\text{SR},i}^2 \sigma_{\text{RD},i}^2 \beta_{\text{RD},k} \left(\left(M + 1\right)\sigma_{\text{SR},i}^2 + {\tilde\sigma}_{\text{SR},i}^2 \right).
\end{align}

e) for $m = i, n =k$, ${\tt E}\left\{|{\bf g}_{\text{RD},k}^T {\bf{\hat G}}_\text{RD}^* {\bf{\hat G}}_\text{SR}^H {\bf g}_{\text{SR},i}|^2 \right\} = 0$.

f) for $m=k, n =i$, ${\tt E}\left\{|{\bf g}_{\text{RD},k}^T {\bf{\hat G}}_\text{RD}^* {\bf{\hat G}}_\text{SR}^H {\bf g}_{\text{SR},i}|^2 \right\} = 0$.

Combining a), b), c), d), e), and f), we have
\begin{align}
  &{\tilde C}_k = M^2 \sum\limits_{i\neq k} \beta_{\text{SR},i} \beta_{\text{RD},k} \sum\limits_{n =1}^K \sigma_{\text{SR},n}^2 \sigma_{\text{RD},n}^2 \\ \notag
  &+ M^3 \sum\limits_{i\neq k}  p_{\text S,i} \left( \sigma_{\text{SR},k}^2 \sigma_{\text{RD},k}^4 \beta_{\text{SR},i} + \sigma_{\text{SR},i}^4 \sigma_{\text{RD},i}^2 \beta_{\text{RD},k}\right).
\end{align}

5) ${\tilde D}_k$:
Following the same approach as with the derivations of ${\tilde B}_k$, we obtain
\begin{align}
  {\tilde D}_k = M^3 \sigma_{\text{SR},k}^2 \sigma_{\text{RD},k}^4 + M^2 \beta_{\text{RD},k} \sum\limits_{n = 1}^K \sigma_{\text{SR},n}^2 \sigma_{\text{RD},n}^2.
\end{align}

6) ${\tilde E}_k$:
By using the fact that ${\tilde E}_k = \left(1 - \frac{2}{\pi}\right)\frac{1}{\alpha_\text a^2} {\tilde D}_k$, we obtain the result for ${\tilde E}_k$.

7) ${\tilde F}_k$:
${\tilde F}_k = \frac{1 - \frac{2}{\pi}}{\alpha_\text a^2 \alpha_\text d^2} {\tt E}\left\{ ||{\bf g}_{\text{RD},k}||^2 \right\} = \left(1 - \frac{2}{\pi}\right)\frac{M \beta_{\text{RD},k}}{\alpha_\text a^2 \alpha_\text d^2}$.

8) ${\tilde G}_k$:
Combining \eqref{eq:gamma}, \eqref{eq:Aa:approx}, and \eqref{eq:Ad:approx}, we can find the value of ${\tilde G}_k$.
\section{Proof of Proposition \ref{prop:rate:compare}}\label{app:prop:rate:compare}
We can readily observe that $R_k^\text{pD} > {\tilde R}_k$ and $R_k^\text p > R_k^\text{pA}$. Thus, we only focus on comparing $R_k^\text{pA}$ and $R_k^\text{pD}$. Due to the fact that ${\hat\sigma}_{\text{SR},k}^2 = \frac{\pi}{2} {\sigma}_{\text{SR},k}^2$ and ${\hat\sigma}_{\text{RD},k}^2 = \frac{\pi}{2} {\sigma}_{\text{RD},k}^2$ (c.f. \eqref{eq:sigma_SR}, \eqref{eq:sigma_RD}, and Corollary \ref{coro:perfect:ADC}), and by neglecting the low order terms as $M \rightarrow \infty$, the ratio between the SINR of $R_k^\text{pA}$ and that of $R_k^\text{pD}$ can be expressed as
\begin{align}
  \frac{2^{\frac{2\tau_\text c}{\tau_\text c - 2\tau_\text p} R_k^\text{pA}} -1}{2^{\frac{2\tau_\text c}{\tau_\text c - 2\tau_\text p} R_k^\text{pD}} - 1} \rightarrow \frac{f_2}{f_1},
\end{align}
where
\begin{align}
f_1 &= \frac{2}{\pi} p_{\text S,k} \sigma_{\text{SR},k}^4 \sigma_{\text{RD},k}^2 \beta_{\text{RD},k}
+ \frac{2}{\pi} p_{\text S,k} \sigma_{\text{RD},k}^4 \sigma_{\text{SR},k}^2 \beta_{\text{SR},k} \\ \notag
&+ \frac{2}{\pi} \sum\limits_{i\neq k}  p_{\text S,i} \left( \sigma_{\text{SR},k}^2 \sigma_{\text{RD},k}^4 \beta_{\text{SR},i} + \sigma_{\text{SR},i}^4 \sigma_{\text{RD},i}^2 \beta_{\text{RD},k} \right) \\  \notag
 &+ \sigma_{\text{SR},k}^2 \sigma_{\text{RD},k}^4 + \left(1 - \frac{2}{\pi}\right)\beta_{\text{RD},k} \sum\limits_{i=1}^K p_{\text S,i}\sigma_{\text{SR},i}^4 \sigma_{\text{RD},i}^2 \\ \notag
 &+\frac{1}{p_\text R}\sum\limits_{i=1}^K p_{\text S,i}\sigma_{\text{SR},i}^4 \sigma_{\text{RD},i}^2 \\
 f_2 &= p_{\text S,k} \sigma_{\text{SR},k}^4 \sigma_{\text{RD},k}^2 \beta_{\text{RD},k}
+ p_{\text S,k} \sigma_{\text{RD},k}^4 \sigma_{\text{SR},k}^2 \beta_{\text{SR},k} \\ \notag
 &+ \sum\limits_{i\neq k}  p_{\text S,i} \left( \sigma_{\text{SR},k}^2 \sigma_{\text{RD},k}^4 \beta_{\text{SR},i} + \sigma_{\text{SR},i}^4 \sigma_{\text{RD},i}^2 \beta_{\text{RD},k} \right)\\ \notag
 &+ \left(\frac{\pi}{2} - 1\right)  \left(1 + \sum\limits_{k=1}^K p_{\text S,k} \beta_{\text{SR},k} \right) \sigma_{\text{SR},k}^2 \sigma_{\text{RD},k}^4 \\ \notag
  &+\frac{1}{p_\text R}\sum\limits_{i=1}^K p_{\text S,i}\sigma_{\text{SR},i}^4 \sigma_{\text{RD},i}^2.
 \end{align}
 Since $f_1 < f_2$, we conclude that $R_k^\text{pA} > R_k^\text{pD}$. This completes the proof.

\bibliographystyle{IEEE}
\begin{footnotesize}
 
 \end{footnotesize}

\end{document}